\documentclass[12pt]{amsart}
\usepackage{amssymb, amsmath}
\usepackage{graphics}
\usepackage{epsfig}
\usepackage{amsxtra}
\usepackage{color}

\numberwithin{equation}{section}

\def\bbbone{{\mathchoice {1\mskip-4mu {\rm{l}}} {1\mskip-4mu {\rm{l}}}
{ 1\mskip-4.5mu {\rm{l}}} { 1\mskip-5mu {\rm{l}}}}}

\newtheorem{thm}{Theorem}[section]
\newtheorem{theorem}{Theorem}
\newtheorem{lemma}[thm]{Lemma}

\newtheorem{prop}[thm]{Proposition}

\theoremstyle{definition}

\newcommand{\defeq}{\stackrel{\rm{def}}{=}}

\renewcommand{\Re}{\operatorname{\rm Re}\nolimits}
\renewcommand{\Im}{\operatorname{\rm Im}\nolimits}

\def \supp {\operatorname{supp}}

\def \Domain {\operatorname{Domain}}
\def \vol {{\rm vol}}
\def \pilambda {\Pi^{\partial X_0}_{\Lambda}}
\def \pilambdanot {\Pi^{\partial X_0}_{\Lambda_0}}
\def \piN {\Pi^{\rm{in}}_{N}}
\def \rsurface {\Lambda_{\sigma(\Delta_{\partial X_0} )}}

\def \rest {\!\!\upharpoonright}
\def \restrict {\!\!\upharpoonright}

\def \mcd {{\mathcal D}}
\def \Real {{\mathbb R}}
\def \RR {{\mathbb R}}

\def \Complex {\mathbb{C}}
\def \Natural {{\mathbb N}}

\def \Integers {{\mathbb Z}}
\def \eff {\operatorname{eff}}

\title [Mahaux-Weidenm\"uller formula]
{A mathematical formulation of the Mahaux-Weidenm\"uller formula for the 
scattering matrix}
   \author { T.J. Christiansen}
\address{Department of Mathematics,
University of Missouri,
Columbia, Missouri 65211, USA} 
\email{christiansent@missouri.edu}
   \author { M. Zworski}  
\address{Mathematics Department, University of
California, Evans Hall, Berkeley, CA 94720, USA}
\email{zworski@math.berkeley.edu}

\begin{document}
\maketitle

\section{Introduction}
\label{int}

The purpose of this paper is
to give a mathematical explanation of a formula for the scattering
matrix for a manifold with infinite cylindrical ends or a waveguide.
This formula, which is well known in the physics literature, is 
sometimes referred to as the Mahaux-Weidenm\"uller formula \cite{MW}. 
We show that a version of this formula given in \eqref{eq:sfinite} below
gives the standard scattering matrix used in the mathematics literature.
We also show that the finite rank approximation of the interaction 
matrix gives an approximation of the scattering matrix with errors
inversely proportional to a fixed dimension-dependent power of the 
rank.



\begin{theorem}
\label{theorem:intro}
Let $ X = X_0 \cup (0, \infty ) \times 
      \partial{X_0} $ be a manifold with cylindrical ends -- see
\S \ref{scat} for a precise definition and 
Figure \ref{f:manifold} for an illustration. Let $ \{\Psi_n\}_{n=0}^\infty$
be an orthonormal set of real
eigenfunctions of the Neumann Laplacian, 
$ -H_{\rm{in}} $, on $X_0$ with eigenvalues $-\tau_n^2$.  Let 
$ \{\varphi_\lambda\} $ be the same set for the Laplacian on $ \partial X_0 $,
with $ -\sigma_\lambda^2 $ denoting the corresponding eigenvalues.
Let us define the interaction matrix  by 
\begin{gather}
\label{eq:W1} 
\begin{gathered}
  W_{N, \Lambda } ( k )  \; : \; L^2 ( \partial X_0 ) 
\longrightarrow L^2 ( X_0 ) \,, \\ W_{N,\Lambda} f  = 
\sum_{ 0\leq \tau_n\leq \sqrt{N}}\Psi_n 
\sum_{0\leq \sigma_\lambda\leq \sqrt{\Lambda}}  
( k^2 - \sigma_\lambda^2)^{\frac14}
\langle \Psi_n\rest_{\partial X_0} , \varphi_{\lambda}\rangle 
 \langle f , \varphi_\lambda 
\rangle \,, \end{gathered}
\end{gather} 
and the effective Hamiltonian by
\[ H_{N,\Lambda} (k) \defeq H_{\rm{in}} - i W_{N, \Lambda } (k) 
 W_{N, \Lambda } (k)^t \,. \]
Then for $k\in \Real$,  
the entries of the scattering matrix (see \S \ref{scat}) are
given by $   S_{ \lambda, \lambda' } ( k ) = \;$
\begin{equation}
\label{eq:main}
- \langle  (I-2 i W_{N,\Lambda} 
(k)^t(k^2-H_{N,\Lambda}(k))^{-1}W_{N,\Lambda}(k)) 
\varphi_{\lambda'},\varphi_\lambda  \rangle  +{\mathcal O} ( N^{-\frac12} +  
e^{- \Lambda/C } ) \,,
 \end{equation}
if $
\sigma_{\lambda}, \sigma
_{\lambda'} \leq |k| $, and 
$ \Lambda > k^2 $. The error bound $ {\mathcal O} ( N^{-\frac12} ) $
is optimal -- see \S \ref{ex}, and the constant can be chosen uniformly for $k$ lying in compact sets.

\end{theorem}
\noindent
Theorem \ref{thm:invertible} provides
 a related result, for other values of $k$.
Also, we remark that the matrix defined by the leading term in \eqref{eq:main},
$\sigma_\lambda, \sigma_{\lambda'} \leq | k | $, 
is in fact unitary -- see Lemma \ref{l:unitary}. 

\medskip

The physics literature contains several versions of the Mahaux-Weidenm\"uller
formula.
One commonly found formula -- see for instance \cite{a-r},\cite{p-s-s}
and references given there -- is given as follows 
\begin{equation}\label{eq:wf1}
\widetilde S_f(k)=-\left(I-2 i W(k)^*(k^2- \widetilde 
H_{\rm{eff}})^{-1}W(k)\right).
\end{equation}
Here
\begin{equation}\label{eq:heff1}
\widetilde H_{\rm{eff}} \defeq H_{\rm{in}}-i  W(k)W(k)^*
\end{equation}
where $-H_{\rm{in}}$ is the Neumann Laplacian 
in the ``interaction region'' $X_0$, a compact piece of the waveguide
or manifold with infinite cylindrical end, and $ W(k) $ is
the frequency dependent interaction matrix.
When applied in numerical simulations only finite number
of modes of $ H_{\rm{in}} $ are taken which results in a finite 
rank approximation of $ W ( k ) $, as described in \eqref{eq:W1}.
The formula, in its finite rank version,
 is the basis of random matrix models in scattering theory --
see \cite[Section III.D]{GWe}. 
For some recent experimental results related 
to the formula see for instance \cite{StK}.

The formula (\ref{eq:wf1}) is not strictly speaking correct. The 
advantage of \eqref{eq:wf1} is that $ \widetilde S_f ( k ) $ is
unitary for real $ k $ by a linear algebra argument. It is also 
close to the correct scattering matrix given below.

As shown in Proposition \ref{prop1}, 
the scattering matrix \cite{tjc} which is standard in the mathematical
literature is  recovered from an expression close to \eqref{eq:wf1}:
\begin{equation}\label{eq:wf}
S_f(k)=-\left(I-2  i W(k)^t(k^2-H_{\rm{eff}})^{-1}W(k) \right)
\end{equation}
with
\begin{equation}\label{eq:heff}
H_{\rm{eff}}=H_{\rm{in}}-i  W(k)W(k)^t \,,
\end{equation}
and, with the notation of \eqref{eq:W1}, 
\[ W ( k ) \defeq W_{\infty, \infty} ( k ) \,.\]
In fact, $-H_{\eff}=-H_{\eff}(k)$ is the Laplacian on $X_0$, with 
a boundary condition that depends on $k$; see Lemma \ref{l:heff}.
Lemma \ref{l:equalresolve} demonstrates the relationship between 
$(k^2-H_{\eff})^{-1}$ and the resolvent of the Laplacian on $X$.

This correct version (\ref{eq:wf})
 appears in \cite{a-r}, though again only a finite 
number of modes are included.
We note that our sign convention, while agreeing with 
\cite{a-r}, is not consistent with many other authors.  It appears that
this sign is correct, and that the difference can be traced to a different
normalization of the scattering matrix.
The difference between (\ref{eq:wf1}) and (\ref{eq:wf}) does
not appear in many of the physics papers, where generally only an 
approximation $W_a(k)$ of $W(k)$ is used, and the approximation is such that
$W_a(k)^*=W_a(k)^t$.  The 
operator $ S_f ( k ) $, unlike $ \widetilde S_f ( k ) $, is typically 
not unitary for real $k$.

However, (\ref{eq:wf}) 
gives what one might call the extended, or full, scattering
matrix.    To get the usual finite
dimensional unitary scattering matrix (whose dimension changes at
roots of the eigenvalues of the cross section of the end), we put, 
for $k$ real,
\begin{equation}\label{eq:sfinite}
 S(k)=-\Pi_{k^2}^{\partial X_0}
\left(I-2  i W(k)^t(k^2-H_{\rm{eff}})^{-1}W(k) \right)\Pi_{k^2}^{\partial X_0}
\end{equation}
where $\Pi_{k^2}^{\partial X_0}$
 projects to the span of the eigenfunctions of $-\Delta_Y$,
with eigenvalue
at most $k^2$. Here $\Delta_Y$ is
the Laplacian on the cross section of the end.  Proposition \ref{prop1} 
shows that this is the unitary scattering matrix which appears
 in the mathematical 
literature.
Lemma \ref{l:unitary} gives an algebraic proof that the matrix given by 
(\ref{eq:sfinite}) is unitary for $k\in \Real$.
  Note that if $k\in \Real$, the operator defined by 
(\ref{eq:wf1}) is
unitary, but the finite rank-operator (corresponding to a 
finite-dimensional matrix)
$$-\Pi_{k^2}^{\partial X_0}
\Big( I-2 i W(k)^*(k^2-H_{\rm{eff}})^{-1}W(k)\Big) 
\Pi_{k^2}^{\partial X_0}$$
with $H_{\rm{eff}}$ given by (\ref{eq:heff1}), is not unitary in general, 
if $ W(k) $ takes into account contributions of evanescent modes.
Evanescent modes correspond to eigenvalues of $ -\Delta_Y $ larger
than $ k^2 $.

Let us add that the articles \cite{a-r} and \cite{p-s-s} already have 
a fairly mathematically careful description of the Mahaux-Weidenm\"uller
formula. In \cite{sss} a detailed analysis of several one dimensional
models is also provided. Another related approach to scattering/transport
is due to Fisher-Lee \cite{FL}, see also \cite{SD}.

\medskip
\noindent
{\sc Acknowledgments.} 
We would like to thank St\'ephane Nonnenmacher for encouraging us
to write this paper, Henning Schomerus for letting us know
about the Fisher-Lee formalism, Ulrich Kuhl for helpful
conversations, and an anonymous referee whose comments 
helped us to clarify the exposition.
Part of the work on this note was done
while the first author was a visitor at MSRI.  
The partial support 
of the work of the first author by MSRI, an MU research leave,
and the NSF grant DMS 0500267 is gratefully acknowledged, as is
that of the second author by 
the NSF grant DMS 0654436.  The first author thanks the Mathematics Department
of U.C. Berkeley for its hospitality in spring 2009.

\medskip
\noindent
{\sc Remark.} We use the notation $( u, v )$ to denote
the Hermitian inner product, and $\langle u, v\rangle$ to denote 
the form which is linear in both arguments.

\section{Scattering matrix}
\label{scat}

In this section we recall the general assumptions for 
manifolds with cylindrical ends and the definition of the
scattering matrix.

\begin{figure}[htp]
\centering
\includegraphics[width=3.5in]{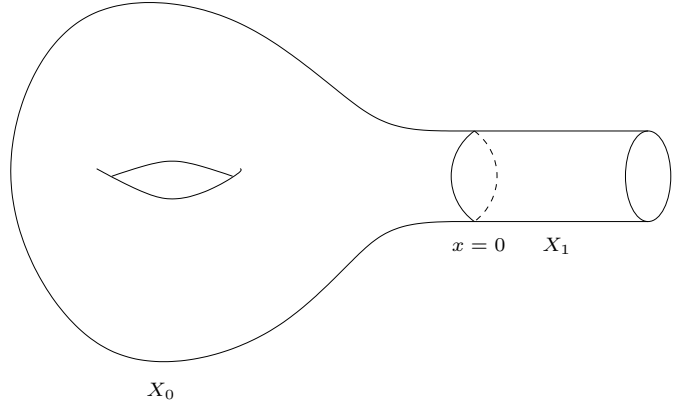}
\caption{An example of a manifold with an infinite cylindrical end.}
\label{f:manifold}
\end{figure}

Our model is a manifold $X$ with infinite cylindrical ends and 
smooth metric $g$-- see Figure \ref{f:manifold}. 
In physics language that means a waveguide with
periodic boundary conditions.
The same arguments apply to waveguides with Dirichlet or Robin 
boundary condition but
we choose to avoid mild technical complications associated with 
that setting.
For purely notational reasons we also assume that there is only one end.
Then 
\[ X=X_0\sqcup (0,\infty)\times Y \,, \ \ Y = \partial X_0 \,, \]
where $ X_0 $ is a compact manifold with a smooth boundary  $Y$.  We require that 
$g\restrict _{[0,\infty)\times Y}=(dx)^2+g_Y$, where $g_Y$ is a metric on $Y$.
Moreover, we choose our decomposition so that there is a neighborhood $U\subset X_0$
of $\partial X_0$ on which $g$ is a also a product: 
\[ g\restrict _U =(dx)^2+g_Y \,. \]


Recall that
$\{\varphi_{\lambda}\}$
are an orthonormal set of eigenfunctions of $\Delta_Y$.
  We use the 
convention that the
 energy is $k^2$, and $k_{\lambda}=\sqrt{k^2-\sigma_{\lambda}^2}$,
with the imaginary part chosen to be non-negative when $\Im k \geq 0$.
We call the region with $\Im k \geq 0$ the physical region.
Given $\lambda \in \Natural$,
if $k$ is in the physical region, and with $ \Im k > 0 $,  there is a unique
$\Phi_{\lambda}(p,k)$
so that
\begin{equation}\label{eq:Phi1}
(-\Delta_X-k^2)\Phi_{\lambda}(p,k)=0\; \text{on $X$} 
\end{equation}
and
\begin{equation} \label{eq:Phi2}
\Phi_{\lambda}\rest_{(0,\infty)\times Y}=
e^{-ik_{\lambda}x}\frac{\varphi_{\lambda}(y)}{\sqrt{k_{\lambda}}}
+ \sum_{\lambda'}S_{\lambda'\lambda}(k) e^{ik_{\lambda'}x}
\frac{\varphi_{\lambda'}(y)}{\sqrt{k_{\lambda'}}}
\end{equation}
for some $S_{\lambda' \lambda}$.
To see this we use the resolvent $ ( - \Delta_X - k^2 )^{-1} $ which
is a bounded operator $ L^2 ( X ) \rightarrow H^2 ( X )$, for $ \Im k> 0$:
\begin{gather*} 
\Phi_{\lambda} ( p , k ) = ( 1 -\psi) \varphi_\lambda ( y ) e^{-ik_\lambda x }
+ ( - \Delta_X - k^2 )^{-1} \left( [\Delta_X , \psi ] 
\left(  \varphi_\lambda ( y ) e^{-ik_\lambda x } \right) \right) (p) \,,\\
\psi \in C^\infty_0 ( X ) \,, \ \ \psi \rest_{X_0} \equiv 1 \,. 
\end{gather*}
Since on $ X_1 $ we have $ -\Delta_X = - \partial_x^2 - \Delta_Y $,
separation of variables shows that $ \Phi_\lambda $ can be written
as in \eqref{eq:Phi2}.

The resolvent, $ ( -\Delta_X - k^2)^{-1} $, 
continues meromorphically to 
\begin{equation}
\label{eq:Riem}  \Lambda_{\sigma(\Delta_{\partial X_0} )} \supset \{ k \; : \;
\Im k > 0  \} \,, 
\end{equation}
a Riemann surface branched at $ \sigma_\lambda $'s 
-- see \cite[Sect.6.7]{Mel}.  We remark that this Riemann surface is such 
that each
$k_{\lambda}$ defined above extends to be a holomorphic single-valued
function.  Thus $ \Phi_\lambda ( p , k ) $ has a meromorphic
continuation to $ \Lambda_{\sigma(\Delta_{\partial X_0} )}$ which
is regular for $ \Im k = 0 $ except when $ k_{\lambda'}$'s are $ 0$, 
or when $ k^2 \in \sigma ( -\Delta_X ) $.

The full, or extended, scattering matrix is the infinite matrix
$$S_f(k)=(S_{\lambda'\lambda}(k))_{\lambda,\lambda'\in \Natural}.$$
For $k\in \Real$, the matrix more commonly called the scattering matrix
is the finite-dimensional matrix
given by
$$S(k)=(S_{\lambda' \lambda}(k))_{\sigma_\lambda^2, \sigma_{\lambda'}^2
\leq k^2}.$$
We remark that if $\Im k>0$, while each entry $S_{\lambda\lambda'}(k)$ 
is
well-defined away from its poles, there is not a canonical choice for 
``the'' scattering matrix.
However, in 
general  it is 
 $(\sqrt{k_\lambda}/ \sqrt{k_{\lambda'}})S_{\lambda'
\lambda}(k)$, not 
$S_{\lambda'\lambda}$, which has a meromorphic continuation to 
$\Lambda_{\sigma(\Delta_{\partial X_0} )}$ for each $\lambda,\;
\lambda'$.
We shall use this continuation in the proof of the theorem.

\section{The formula}
\label{form}

Let $\Delta_Y$ be the Laplacian on $Y$, and let
$\{\sigma_{\lambda}^2\}$ be the eigenvalues of $-\Delta_Y$, repeated 
according to multiplicity,
and 
let $\{\varphi_{\lambda}\}$ be an associated set of real, orthonormal 
eigenfunctions of the Laplacian on $Y$.  
Let
$-H_{\rm{in}}$ be the Laplacian with 
Neumann boundary conditions on $X_0^{\circ}$,
and let  $\{\Psi_n\}$ be a
set of real, orthonormal eigenfunctions 
of $H_{\rm{in}}$.


First, we define the operator $W(k)$ by explicitly giving its Schwartz 
kernel.  Our starting point is the representation of $W(k)$ from 
\cite{a-r} or \cite{p-s-s}.
  We write $p$ to represent a point in ${X_0}$, and 
$y$ or $y'$ to represent a point in $Y$; on  $U\subset X_0$ we 
may write $p=(x,y)$, with $\{x=0\}= \partial X_0$. 
Then, with 
\[ \Psi_{n,\lambda}(0) \defeq \int_Y\varphi_{\lambda}(y)\Psi_n(0,y) \,, \]
we follow the physics literature and 
define the {\em coupling operator} by giving its integral kernel 
(with integration with respect to Riemannian densities) as 
\begin{equation}
\label{eq:W}
\begin{split}
W(p,y')& \defeq \sum_{n,\lambda}\sqrt{{k_{\lambda}}}
\Psi_{n}(p)\Psi_{n,\lambda}(0)\varphi_\lambda(y')  \\
& = \sum_{n} 
\Psi_{n}(p)P_k\Psi_{n}(0,y').
\end{split}
\end{equation}
Here $P_k=(k^2+\Delta_Y)^{1/4}$ is defined by
$P_k\varphi_{\lambda}= \sqrt{k_{\lambda}}\varphi_{\lambda}$.  While either choice of 
the square root is possible, it is crucial that this is consistent with
that used to define the scattering matrix; see (\ref{eq:Phi2}). 
The series converges in the sense of distributions. Hence,  $ W ( p, y') $
is understood as a distribution on $ X \times Y $ -- see Lemma \ref{l:0}
below.

This definition (\ref{eq:W})  appears normally in the physics literature.
When we take {\em all} the eigenstates then $ W $ (and especially 
$ W^t $) takes a very simple form given in the following lemma.
When, as is done in the physics literature, we take only finitely 
many states, the formula for $ W^t $ is given in the remark after
the lemma.

Let $ \dot{\mathcal D}' ( X_0 ) $ be the space of distributions
on $ X $ supported in the (closed) set $ X_0 $. 
We also use the following convention: for $ s\geq 0 $ 
the space 
$ H^s ( X_0 ) $ denotes restrictions of elements of $ H^s ( X ) $ to 
$ X_0 $, while for $ s \leq 0 $, $ H^s ( X_0 ) $ denotes 
elements of $ H^s (X ) $ supported in the (closed) set $ X_0 $.
See 
\cite[Appendix B.2]{Hor} for a careful discussion: in the notation 
used there
\[ H^s ( X_0 ) = \left\{\begin{array}{ll}
\bar{H}_{(s)} ( X_0 ) & s \geq 0 \\
\dot{H}_{(s)} ( X_0 ) & s \leq 0 \,.
\end{array} \right.\]
With this
notation in place we can formulate
\begin{lemma}
\label{l:0} 
The operator \eqref{eq:W} is equal to 
\begin{equation}
\label{eq:l01} 
   W(k) g = \delta_{\partial X_0} \, P_{k} g \,, \ \ g \in 
C^\infty (\partial  X_0 ) \,, 
\end{equation}
where $ \delta_{ \partial X_0} \in \dot{\mathcal D}' ( X_0 ) $ is the 
distribution defined by
\[  \delta_{ \partial X_0} ( \varphi ) = \int_{\partial X_0} \varphi
\rest_{\partial X_0} d {\rm{vol}}_Y \,. \]
We have 
\begin{equation}
\label{eq:l03} 
W(k):H^{s}(Y)\rightarrow H^{\min(-1/2-,s-1)}(X_0)\,, \ \ s \in \RR  \,, 
\end{equation}
and $ W(k)g\rest_{X_0^\circ}=0 $.
The transpose, $W(k)^t : H^{1/2 + s } (X_0 ) \rightarrow 
H^{-1/2 + s } ( Y )  $, $ s > 0 $, is given by 
\begin{equation}
\label{eq:l02}   W(k)^t f ( y ) = P_k ( f\rest_{\partial X_0} ) \,. 
\end{equation}
\end{lemma}
\begin{proof}
To prove \eqref{eq:l01} we need to compute, in the 
notation of distributions,  $  W ( f \otimes g ) $, where 
$ f \in \bar C^\infty_0 ( X_0 ) $. The definition \eqref{eq:W}
gives
\[  \begin{split}
W( f \otimes g ) & = \sum_{n} \left( \int_{X_0}  \Psi_n f  \right)
\left( \int_{ \partial X_0 } \Psi_n \rest_{ \partial X_0} P_k g \right)  \\
& = \int_{ \partial X_0 } \left( \sum_{n} 
\left( \int_{ X_0 }  \Psi_n f \right) \Psi_n \rest_{ \partial X_0} \right) 
P_k g  \\
& = \int_{ \partial X_0 }  f\rest_ {\partial X_0} P_k g  \,,
\end{split} 
\]
which proves \eqref{eq:l01} and, by duality, \eqref{eq:l02}. The 
mapping property of $ W ( k )^t $ follows from the fact that
$ f \mapsto f \rest_{ X_0} $ takes $  H^{s+1/2} ( X_0 ) $ to 
$ H^s (\partial X_0 )  $ for $ s > 0 $, and $ P_k : H^s ( \partial X_0 ) 
\rightarrow H^{s-1/2} ( \partial X_0 ) $. The mapping property 
\eqref{eq:l03} follows by duality.
\end{proof}

\medskip

\noindent
{\bf Remark.} In Lemma \ref{l:0} all the structure of the the basis
of eigenvectors of $ H_{\rm{in}} $ and $ \Delta_Y $ disappears. The question
which we address in Section \ref{s:accuracy}
is how close the approximation based on using only 
finitely many basis elements gets to the actual scattering matrix.
Then for $ a = ( N, \Lambda ) \in [0, \infty]^2 $ we define 
\[  W_a ( k )^t \defeq P_k \, \bbbone_{[0,N] } ( \Delta_Y ) R \, 
\bbbone_{[0,\Lambda] } 
( H_{\rm{\rm{in}}} ) \,, \ \  R u \defeq u \rest_{\partial X_0} \,. \]
We note that 
\[ W_{( \infty, \infty ) }( k ) = W ( k ) \,,\]
and that for $ N < \infty $ and $ \Lambda < \infty$,
\[ W_a ( k ) \; : \;  {\mathcal D}' ( Y ) \longrightarrow 
C^\infty ( X_0 ) \,, \ \ 
 W_a ( k )^t \; : \;  \dot {\mathcal D}' ( X_0 ) \longrightarrow 
C^\infty ( Y ) \,,\]
where $ C^\infty ( X_0) $ denotes {\em extendable} smooth functions on the
compact manifold $ X_0 $.


\medskip

We make the definition (\ref{eq:heff}) 
of $H_{\eff}$ rigorous via the quadratic form
\begin{align*}q(u,v)& =q(k)(u,v) = 
\int_{X_0} \nabla u \overline{\nabla v}\; d\vol_{X_0}
-i \int_{\partial X_0} W^t(k) u \overline{W^*(k) v}\; d\vol_Y\\ &
= \int_{X_0} \nabla u \overline{\nabla v}\; d\vol_{X_0}
-i \int_{\partial X_0} P_k R u \overline{P_k^* R v}\; d\vol_Y
\end{align*}
with form domain $H^1(X_0)$.
If 
$$q(u,v)=(w,v)$$
for some $w\in L^2(X_0)$ and all $v\in H^1(X_0)$, then $u$ is in the domain of 
$H_{\eff}$ and 
$H_{\eff}u=w.$
Moreover, 
$$(w,v)=q(u,v)=-\int_{X_0}\Delta_{X_0}u \overline{v}+ \int_{\partial X_0}
(\partial_n u-i P_k^2 R u)\overline{v}\,,$$
where $ \partial_n u $ denotes the outward unit normal derivative at the
boundary. 
Since this must hold for all $v\in H^1(X_0)$, $-\Delta_{X_0}u=w$ and 
$$0= \partial_n u-iP^2_kR u.$$ We note that $ u \in H^2 ( X_0 ) $ where
the space is defined by restricting elements of $ H^2 ( X ) $ to $ X_0 $
-- see \cite[Appendix B]{Hor}. 
We summarize this in the following
\begin{lemma}\label{l:heff}
Suppose $ u \in   \Domain(H_{\eff})$.  Then 
$u\in H^2(X_0)$, and
\[ H_{\eff} u= -\Delta_{X_0}u \,,  \ \ \partial_n u- iP_k^2 Ru=0 \,. \]
\end{lemma}

Next we investigate the relation between $(k^2-H_{\eff})^{-1}$ and the resolvent of the 
Laplacian on $X$.
Denote
$$R_X(k)\defeq (k^2+\Delta_X)^{-1},\;\text {for $\Im k >0$}.$$
Then, for $K\subset X$ any compact set  $\bbbone_K R_X(k)\bbbone_K$ 
has a meromorphic extension to 
$ \Lambda_{\sigma(\Delta_{\partial X_0} )} $, see \cite{Mel}.  
In Lemma \ref{l:mero}
 we shall show that $(k^2-H_{\eff})^{-1}:L^2(X_0)\rightarrow H^2(X_0)$
exists for $k^2 \ll 0$, $\Im k>0$, and is meromorphic on
  $ \Lambda_{\sigma(\Delta_{\partial X_0} )} $, the Riemann
surface \eqref{eq:Riem}.  One could provide an alternate 
proof using the first part of the proof of  Lemma \ref {l:equalresolve}  and the results of 
\cite{Mel} on the meromorphic continuation of $R_X(k)$.

We remark that when we use $k\in \Lambda_{\sigma(\Delta_{\partial X_0} )}$, by
abuse of notation we mean by $k^2$ the complex number which is the 
continuation of $k^2$ from the physical half plane $\Im k \geq 0$.

\begin{lemma}\label{l:equalresolve}
We have the following relation between $R_X(k)$ as defined above and
$(k^2-H_{\eff})^{-1}= (k^2-H_{\eff}(k))^{-1}$:
$$(k^2-H_{\eff})^{-1}= \bbbone_{X_0} R_X(k) \bbbone_{X_0}$$
for $k\in \rsurface$.
In particular, the poles of $(k^2-H_{\eff})^{-1}$ are the same as the 
poles of $R_X(k)$.
\end{lemma}
\begin{proof}
Suppose $g\in L^2(X_0)\subset L^2(X)$ and 
$g$ is $0$ in a neighborhood of $\partial X_0$.
Then
\begin{equation}\label{eq:good}
(k^2+\Delta_{X_0})\bbbone_{X_0} R_X(k) g = 
(k^2+\Delta_{X_0})\bbbone_{X_0} R_X(k)\bbbone_{X_0} g= g \;
 \text{ on $X_0^{\circ}$}
\end{equation}
for $k\in 
\Lambda_{\sigma(\Delta_{\partial X_0} )} .$

Note that since $\supp g \subset X_0$, $(k^2+\Delta_X)g=0 $ on $X_1$.
Then for  $\Im k>0$ (that is, for $k$ in the physical space),
the requirement that
$R_X(k)g\in L^2(X)$ means that
\begin{equation}\label{eq:end}
R_X(k)g\restrict_{X_1}= \sum a_\lambda e^{ik_\lambda x} \varphi_\lambda
\end{equation}
for some constants $a_\lambda= a_\lambda(k)$.  
But then, using the support conditions of $g$
there is a neighborhood
$\tilde{U}\subset X_0$ of $\partial X_0$ so that 
$$R_X(k) g\restrict_{\tilde{U}}= 
\sum a_\lambda e^{ik_\lambda x} \varphi_\lambda.$$
Thus
$$(\partial _n -iP_k^2) (R_X(k)g\restrict_{X_0})\restrict_{\partial X_0} =0$$
so that
$R_X(k)g\restrict_{X_0} $ is in the domain of $H_{\eff}=H_{\eff}(k)$.
Together with (\ref{eq:good}), this means that 
$$(k^2-H_{\eff})^{-1}g= \bbbone_{X_0}R_X(k)g$$
for all $k$ with $\Im k>0$ and all $g\in L^2(X_0)$ which are $0$ 
in a neighborhood of $\partial X_0$.
Since such $g$ are dense in $L^2(X_0)$, this must in fact hold for all
$g\in L^2(X_0)$. 

Since $(k^2-H_{\eff})^{-1}= \bbbone_{X_0}R_X(k)\bbbone_{X_0}$ for all $k$
with $\Im k>0$ and since both sides have meromorphic continuations to 
$ \Lambda_{\sigma(\Delta_{\partial X_0} )} $ (see \cite{Mel} and 
Lemma \ref{l:mero}), they must in fact agree
for all $k\in \Lambda_{\sigma(\Delta_{\partial X_0} )} $.  
\end{proof}

\begin{lemma}   Suppose $(k^2-H_{\eff})^{-1}$ exists. 
For $f\in H^{1} (\partial X_0)$, let 
\[ u= (k^2-H_{\eff})^{-1} W(k) f \,. \]
Then
\begin{equation}
\label{eq:le23}
(k^2 +\Delta_{X_0})u  = 0 \; \text{on}\; X_0^{\circ} \,, \ \ 
(\partial_n -iP^2_k)u \restrict_{\partial X_0} = -P_k f.
\end{equation}
\label{l:heffW}
\end{lemma}
\begin{proof}
We first claim that there exists $F\in H^2(X_0)$ such that 
\begin{equation}
\label{eq:defft} (\partial_{n}-iP_k^2)F\restrict_{\partial X_0} =-P_k f\,.
\end{equation}
In fact, for $ 0 < h \ll 1 $ define $ N ( h ) :  
H^{s} ( \partial X_0 ) \rightarrow H^{s-1} ( \partial
X_0 ) $ as follows:
\[  N(h) = \frac 1 h  \langle I - \Delta_{\partial X_0} \rangle^{\frac12} \,, \ \ \ 
N(h)^{-1} = {\mathcal O} ( h )  \; : \; H^{s-1} ( \partial X_0 ) \longrightarrow 
H^s ( \partial X_0 ) \,. \]
Let $ \chi \in C^\infty_0 ( [ 0 , \epsilon)  ) $ be equal to $ 1 $ in a small 
neighbourhood of $ 0 $, with $ \epsilon $ chosen so that 
$ X_0 \simeq ( - \epsilon , 0 ]_x \times \partial X_0 $, near the boundary.
We define (note that $ x < 0 $)
\begin{gather*} [ T ( h ) g ] ( x, y )  \defeq 
\chi ( - x ) \exp ( x \langle I - \Delta_{\partial X_0} \rangle^{\frac12} 
/ h )  g (y ) \,, \\ 
T( h ) \; : \; H^{s} ( \partial X_0 ) \rightarrow H^{s+ \frac12} ( X_0 ) \,, \ \
s \geq 0 \,, \end{gather*}
so that 
\[  T ( h )  g \rest_{\partial X_0 } = g \,, \ \ 
\partial_n T ( h )  g \rest_{ \partial X_0 } = N ( h ) g \,. \]
For a fixed $ k $, $ P_k^2 = {\mathcal O} ( 1 ) : H^{3/2} ( \partial X_0 ) 
\rightarrow H^{1/2} ( \partial X_0 ) $, and hence, if $ h $ is small enough,
we have the following inverse
\[ ( N ( h ) - i P_k ^2 )^{-1} = N(h)^{-1} ( I - i P_k^2 N(h)^{-1} )^{-1} 
\; : \; H^{\frac12} ( \partial X_0  ) \longrightarrow H^{\frac32} ( \partial X_0 ) \, . \]
Using this and the mapping properties of $ T ( h ) $ we 
construct
\[ F \defeq - T ( h ) ( N ( h ) - i P_k^2 )^{-1} P_k f \in H^2 ( X_0 ) \,, \]
which satisfies \eqref{eq:defft}.

We now set
$$v=F - (k^2-H_{\eff})^{-1} (k^2+\Delta_{X_0})F \,, $$
and observe that 
$v$ satisfies the equations (\ref{eq:le23}).  It remains to show
that $v=u$.

To see that we let $h\in C^\infty (X_0)$,  and apply Green's formula to 
compute
\begin{align*}
((k^2-H_{\eff})^{-1}(k^2+\Delta_{X_0})F, h) & = 
((k^2+\Delta_{X_0})F,((k^2-H_{\eff})^{-1})^* h)\\
 & = \int_{\partial X_0}
\left( \partial_{n}F \overline {((k^2-H_{\eff})^{-1})^* h}
- F \overline{\partial_n((k^2-H_{\eff})^{-1})^* h}\right) \\
& \ \ \ + 
(F,(k^2+\Delta_{X_0}) ((k^2-H_{\eff})^{-1})^* h)
 \\
& =  \int_{\partial X_0} \left( \partial_{n}F \overline {((k^2-H_{\eff})^{-1})^* h}
- F \overline{\partial_n((k^2-H_{\eff})^{-1})^* h}\right) \\
& \ \ \ + ( F , h ) \,. 
\end{align*}
Now we use that 
$\partial_{n}F\rest_{\partial X_0 }  = iP_k^2( F\rest_{\partial X_0})  -P_k f$,
and that 
\[ w\in H^2(X_0) \cap \Domain((k^2-H_{\eff})^*) \  \Longrightarrow \ 
\partial_n w\rest_{\partial X_0} +i(P_k^2)^*Rv=0 \,. \]
Thus we have 
\begin{align*}& 
 ((k^2-H_{\eff})^{-1}(k^2+\Delta_{X_0})F, h)\\
 & = 
\int_{\partial X_0} \left(
(iP^2_k ( F\rest_{\partial X_0})  -P_kf)\overline{((k^2-H_{\eff})^{-1})^* h}
- iF \overline{((P_k^2)^*(k^2-H_{\eff})^{-1})^* h}\right)
+(F, h)\\
 & = - \int_{\partial X_0} P_k f
\overline{((k^2-H_{\eff})^{-1})^* h}  +(F, h) = 
 \int_{X_0 }\left( - ( k^2 - H_{\eff } )^{-1} \delta_{X_0 } P_k f + F
\right) \bar h \,,
\end{align*}
where the last expression follows from the 
definition of $ \delta_{\partial {X_0 }}$.
Since this holds for 
all $h\in C^\infty (X_0)$,  
\[ 
v = F - (k^2-H_{\eff})^{-1}(k^2+\Delta_{X_0})F=
(k^2-H_{\eff})^{-1} \delta_{X_0} P_kf = u \,, \]
proving the lemma.
\end{proof} 

We can now state and prove the main result of this section.
It provides a justification of \eqref{eq:wf} and 
\eqref{eq:sfinite}.

\begin{prop}
\label{prop1}
Let $ W $ be given by \eqref{eq:W}. Then the $\lambda \lambda'$ 
entry of the scattering matrix
defined in \S \ref{scat} is given by 
\begin{equation}
\label{eq:prop1}
 S_{\lambda, \lambda'} ( k ) = \langle S_f ( k ) \varphi_\lambda , 
\varphi_{\lambda' } \rangle_{L^2 ( \partial X_0 ) }  \,, 
\end{equation}
where
\[ S_f(k)=-\left(I-2  i W(k)^t(k^2-H_{\rm{eff}})^{-1}W(k) \right) \,,\]
and $ H_{\rm{eff}} $ is defined in Lemma \ref{l:heff}.
\end{prop}
\begin{proof}
We use Lemma \ref{l:heffW} to express the action of 
$(k^2-H_{\eff})^{-1} W(k) $.
Suppose 
$v_{\lambda}= (k^2-H_{\eff})^{-1}W(k)\varphi_{\lambda}.$
Let $U\subset X_0$ be a neighborhood of $ \partial X_0$.  On $U$
we may use coordinates
 $(x,y)$, with $y\in Y$.  Since
$v_{\lambda}$ lies in the null space of $-\Delta_{X_0} -k^2$, we have that 
$$v_{\lambda}\rest_{U}=\sum_{\lambda'}(a_{\lambda'}e^{ik_{\lambda'}x} + 
b_{\lambda'}e^{-ik_{\lambda'}x}) \varphi_{\lambda'}(y).$$
The boundary conditions \eqref{eq:le23} 
applied to $v_{\lambda}$ at $ \partial X_0$ 
mean that 
$$\sum_{\lambda'} ik_{\lambda'}(a_{\lambda'}-b_{\lambda'})\varphi_{\lambda'}
-i\sum _{\lambda'} k_{\lambda'}(a_{\lambda'}+b_{\lambda'})\varphi_{\lambda'}
= -P_k\varphi_{\lambda}.$$
Then $b_{\lambda}= 1/({2 i} {\sqrt{k_{\lambda}}}) 
$ and $b_{\lambda'}=0$ if $\lambda'\not = \lambda$.  Thus $ v_{\lambda}$ 
is the restriction to $X_0$ of ${-i} \Phi_{\lambda}/2$,
where $\Phi_{\lambda}$ is
determined by (\ref{eq:Phi1}) and (\ref{eq:Phi2}): 
\begin{equation}\label{eq:psilambda}
\Phi_{\lambda}\rest_{(0,\infty)\times Y}=
e^{-ik_{\lambda}x}\frac{\varphi_{\lambda}(y)}{\sqrt{k_{\lambda}}}
+ \sum_{\lambda'}S_{\lambda'\lambda} e^{ik_{\lambda'}x}
\frac{\varphi_{\lambda'}(y)}{\sqrt{k_{\lambda'}}} \,.
\end{equation}
Therefore
\begin{align*}
W(k)^t v_{\lambda}& = 
\sum_{\lambda'}\sqrt{k_{\lambda'}}a_{\lambda'}
\varphi_{\lambda'}
- \frac{i}{2  }\varphi_{\lambda}
 =- \frac{i}{2 }\left(
\sum_{\lambda'}S_{\lambda'\lambda} 
\varphi_{\lambda'}
+ \varphi_{\lambda}\right) \,, 
\end{align*}
which proves the proposition.
\end{proof}

The equation \eqref{eq:prop1} is valid for all real  values of 
$k $ (that is, $k$ on the boundary of 
the physical space) with $k^2>\sigma_\lambda^2,\; \sigma_{\lambda'}^2$, 
 since the matrix coming from the right hand side is unitary and hence the 
singularities of $\langle  S_f(k)\varphi_{\lambda'}, \varphi_\lambda \rangle
_{L^2(\partial X_0)}$ resulting from poles of
$ ( H_{\rm{eff}} - k^2)^{-1} $ are removable.

\section{Accuracy of Approximations}\label{s:accuracy}
Here we investigate the accuracy of the approximations made to use
 (\ref{eq:wf}) in numerical computations.  
Set
\begin{align*} \pilambda f  & \defeq
 \bbbone_{[0,\Lambda]}(-\Delta_{\partial X_0}) f  \ \ \text{for $f\in
   L^2(\partial X_0)$,} \\  \piN g  & \defeq\
\bbbone_{[0,N]}(H_{{\rm in}}) g \ \ \text{for $g\in
   L^2( X_0)$}.
\end{align*}
In parallel with this, we introduce
\begin{align*}
W_{\infty,\infty}(k)&\defeq  W(k) \,,  \ \ \ 
W_{\infty,\Lambda}(k)  \defeq W \pilambda\,, \\
W_{N,\Lambda}(k)& \defeq \piN W(k) \pilambda = \piN W_{\infty,\Lambda} (k)\,,
\end{align*}           
and 
\begin{align*}
H_{\infty,\infty}&\defeq H_{\eff} \,, \ \ \ 
H_{N,\Lambda} \defeq
H_{in} - iW_{N,\Lambda}W_{N,\Lambda}^t, \; N \in \Real \cup \{\infty\} \,. 
\end{align*}
Although $W_{N,\Lambda}$, $W_{\infty,\Lambda}$ depend on $k$, 
for simplicity we generally omit this in our notation.  Note that $H_{\eff}$,
$H_{\infty, \Lambda}$ and $H_{N,\Lambda}$ also depend on $k$.
A quadratic form argument 
(see Lemmas \ref{l:heff} and \ref{l:resN}), 
using the form domain
$H^1(X_0)$, shows that if $u$ is in the domain
of $H_{\infty,\Lambda}$, then $\partial_{n}u-iP_k^2\pilambda R u=0$.
However,
for $N<\infty$ the domain of $H_{N,\Lambda}$ is the set of elements of
$H^2(X_0)$ which satisfy the Neumann boundary condition, $ \partial_n u = 0 $.

Likewise, we define the approximations of the (full) scattering matrix 
obtained by using the approximation $H_{N,\Lambda}$
of $H_{\eff}$ by $S_{f,N,\Lambda}$:
\begin{equation}\label{eq:sfnlambda}
 S_{f,N,\Lambda} (k)=-\Big(I-2  i W_{N,\Lambda} (k)^t(k^2-H_{N, \Lambda})^{-1}
W_{N,\Lambda} (k) \Big) \,. 
\end{equation}
In order to bound the error in these approximations, we shall first
see how close $\Pi^{\partial X_0}_{\Lambda_0} S_{f,\infty,\Lambda}$ is to 
$\Pi^{\partial X_0}_{\Lambda_0} S_{f,\infty,\infty}$, and
then study the difference 
\[ \Pi^{\partial X_0}_{\Lambda_0}
 \left(S_{f,\infty,\Lambda}-S_{f,N,\Lambda}\right) 
\Pi^{\partial X_0}_{\Lambda_0}\,. \]

\subsection{Projection on $\partial X_0$}
We first analyze the approximation with a finite $ \Lambda $ and
$ N = \infty$. The spectral cutoff for the boundary Laplacian, 
$ \Lambda $ has to be taken large enough to guarantee that 
$ \Im k_\lambda > 0 $ for $ \sigma_\lambda^2 > \Lambda $. The
errors then come from {\em evanescent modes} and can be 
estimated using exponential decay. We present the results
in two lemmas.

  Recall
that $H_{\eff}=H_{\eff}(k)$ is defined for $k\in 
\Lambda_{\sigma(\Delta_{\partial X_0} )}$.
\begin{lemma}\label{l:mero}
Let $ \Lambda_{\sigma(\Delta_{\partial X_0} )} $ be the Riemann surface,
given in \eqref{eq:Riem}, 
to which the resolvent of $ -\Delta_X $, $ (-\Delta_X - k^2)^{-1}, $
has a meromorphic continuation (see \cite[Sect.6.7]{Mel}). Then the
operators
$(k^2-H_{\rm{eff}})^{-1}$ and $(k^2-H_{\infty,\Lambda})^{-1}$ are 
meromorphic on  $ \Lambda_{\sigma(\Delta_{\partial X_0} )}
$.
If $k^2-H_{\eff}=k^2-H_{\eff}(k)$ is invertible, so is
$k^2-H_{\infty,\Lambda}(k)$ for $\Lambda > \Lambda_0 ( k_0 ) $ 
sufficiently large,
and $$\|(k^2-H_{\rm{eff}})^{-1}-(k^2-H_{\infty,\Lambda})^{-1}\|_{L^2\rightarrow 
L^2} \leq C \Lambda^{-1/2}\,, \ \ \Lambda > \Lambda_0 ( k ) \,. $$
Moreover, for $k$ restricted
to a  compact set $K\subset \Lambda_{\sigma(\Delta_{\partial X_0} )} $ on on which $k^2-H_{\eff}$ is invertible, $\Lambda_0$ and $C$ can be 
chosen independently of $k$.
\end{lemma}

\begin{proof}
Recall that $U$ is a neighborhood of $\partial X_0$ which we may
identify with $(-\epsilon, 0]_x\times Y$ with  $g\restrict U
= (dx)^2+g_Y$.  Choose $\chi_i\in C^{\infty}(X)$, $i=1,2$, so that
each $\chi_i$ has support in $U$, $\chi_i=1$ in a smaller neighbourhood
of the boundary, and 
$$\chi_{1}\chi_2 =\chi_1 \,, \  \  \supp \chi_2' \cap \supp \chi_1 = 
\emptyset \,. $$
  Set $R_{\Lambda,
  e}(k)$
to be the operator on $L^2 ( (-\infty, 0]\times Y)$ defined by the
Schwartz kernel 
$$ R_{\Lambda, e} ( k )  ( x, y; x',y') 
\defeq \sum_\lambda \frac{1}{2ik_{\lambda}}
(e^{ik_{\lambda}|x-x'|}+ (1-\pilambda)e^{ik_{\lambda}|x+x'|}) 
\varphi_{\lambda}(y)\varphi_{\lambda}(y').
$$  
Note that $R_{\Lambda,e}(k)$ is a meromorphic function of $k\in \Lambda_{\sigma(
\Delta_{\partial X_0})}$ since $k_{\lambda}$ is holomorphic 
on $\Lambda_{\sigma(
\Delta_{\partial X_0})}$.
Let $R_{\infty, e}(k)$ be the operator with Schwartz kernel given by
$$ \sum_\lambda \frac{1}{2ik_{\lambda}}
e^{ik_{\lambda}|x-x'|} 
\varphi_{\lambda}(y)\varphi_{\lambda}(y') \,, 
$$  
and set
$$ E_{\Lambda}(k)= (1-\chi_1) (k^2-H_{\rm{in}})^{-1} 
+ \chi_2 R_{\Lambda, e} ( k ) \chi_1\,, \ \
\Lambda \in \Real \cup \{\infty\} \,.  $$ 
Then, for the same values of $\Lambda$, $E_{\Lambda} v $ satisfies the
boundary
conditions of $H_{\infty, \Lambda}$,  that is 
\[  ( \partial_{n} -iP_k^2\pilambda ) E_\Lambda v \rest_{\partial X_0} =0
\,, \]
and is meromorphic on 
$ \Lambda_{\sigma(\Delta_{\partial X_0} )} $.  Moreover,
\begin{align*}
(k^2+\Delta_{X_0})E_{\Lambda}(k)& = 
I-[\Delta_{X_0},\chi_1]  (k^2-H_{\rm{in}})^{-1}
 + [\Delta_{X_0},\chi_2]R_{\Lambda,e}(k)\chi_1  \\
& \defeq I +K_{\Lambda}(k)
\end{align*}
where $K_{\Lambda }(k)$ is a compact operator.  Moreover,
$K_{\Lambda}(k)$
is a meromorphic function of $k$ in $ \Lambda_{\sigma(\Delta_{\partial
    X_0} )} $
with finite-rank poles.
When $k\in i\Real_+$, $\| K_{\Lambda}(k)\| \rightarrow  0$
as $k^2 \rightarrow -\infty$.   Thus
  $I+K_{\Lambda}(k)$ is invertible for
$k\in i \Real_+$, $-k^2 \gg 0$, and by analytic Fredholm theory (see
for instance \cite[\S 2.4]{SZ})
we have that 
$$(k^2 - H_{\infty, \Lambda}(k))^{-1}=
E_{\Lambda}(k)(I+K_{\Lambda}(k))^{-1}$$
for $k$ in the physical space, and it has a 
 meromorphic continuation to $ \Lambda_{\sigma(\Delta_{\partial X_0} )} $.

Now 
$$\|E_{\Lambda}(k)-E_{\infty}(k)\|_{L^2\rightarrow 
L^2}\leq C \max_{\sigma^2_\lambda >
  \Lambda}
|k_{\lambda}|^{-1}.
$$ 
For $k$ in a compact set of 
$\Lambda_{\sigma(\Delta_{\partial X_0})}$
and $ \sigma_\lambda >\Lambda \geq \Lambda_0(k) $, sufficiently large, 
we have $\Im k_{\lambda} > 0$, and since 
\[ x \in  \supp \chi'_2 \,, \ \  x' \in \supp \chi_1  
\ \Longrightarrow \ | x - x'| , |x+x'| > \epsilon_0 > 0 \,, \]
we have 
$$\| K_{\Lambda}(k)-K_{\infty}(k)\|_{L^2\rightarrow 
L^2} \leq C 
\max _{\sigma^2_\lambda >\Lambda } \frac{|k_{\lambda}|+1
}{|k_\lambda|} e^{-\epsilon_0 \Im k_{\lambda}/2}.$$
This constant is independent of $k$.
Thus, if $\Lambda$ is big enough, $I+ K_{\Lambda}(k)-K_{\infty}(k)$ is
invertible with small norm, and 
$$\| (k^2-H_{\rm{eff}})^{-1}- (k^2-H_{\infty,\Lambda})^{-1}\| 
\leq C \Lambda^{-1/2}
$$
for $\Lambda$ sufficiently large (depending on $k$ or $K$, $\epsilon_0 $ and 
$\| (k^2 -H_{\eff})\|^{-1}$).  The constant can be chosen independently of $k$
on a fixed compact set $K$ where $k^2-H_{\eff}$ is invertible.
\end{proof}

\medskip

\noindent
{\bf Remark.} 
Using this Lemma and the definition (\ref{eq:sfnlambda}) of 
$S_{f,\infty ,\Lambda}$,
 we can see that for $\Lambda \in \Real_+
\cup \{\infty\}$,
\begin{equation}\label{eq:scont}
P_k^{-1}S_{f,\infty,\Lambda}(k)P_k
= 
-(I
-2iP_k^{-1}W_{\infty,\Lambda}(k)^t(k^2-H_{\infty,\Lambda})^{-1}W_{\infty,\Lambda}(k)P_k)
\end{equation}
has a meromorphic continuation to $\Lambda_{\sigma(\Delta_{\partial
    X_0} )}$.  The conjugation by $P_k$ is necessary because while
$P_k^2$
is a well-defined operator for $k\in \Lambda_{\sigma(\Delta_{\partial
    X_0} )}$, 
$P_k$ is not.  Thus the operators $W_{\infty,\Lambda}(k)P_k$ and $P_k^{-1}W_{\infty,\Lambda}(k)^t$
are well-defined on $ \Lambda_{\sigma(\Delta_{\partial
    X_0} )}$, while in general 
$W_{\infty,\Lambda}(k)$ and $W_{\infty,\Lambda}^{t}(k)$ are not.
The existence of the meromorphic continuation of 
(\ref{eq:scont}) means that
$(\sqrt{k_{\lambda'}}/\sqrt{k_{\lambda}})
\langle  S_{f,\infty,\Lambda} (k) \varphi_{\lambda'}, \varphi_\lambda \rangle$
has a meromorphic continuation to $\Lambda_{\sigma(\Delta_{\partial X_0} )}$.

\medskip

\begin{lemma} 
\label{l:23}
Fix $\Lambda_0<\infty$ and $k$ so that $k^2-H_{\eff}$ is
  invertible
and $\Im k_{\lambda}>0$ if $\sigma^2_\lambda >\Lambda_0$.
Suppose $f\in L^2(\partial X_0)$ satisfies 
$\pilambda f= f$ for $\Lambda \geq \Lambda_0$.
  Then, for $\Lambda\geq\Lambda_0$ such that $k^2-H_{\infty,\Lambda}$ is
invertible, we have for some $\epsilon' >0$
$$\| P_k^{-1}\Pi_{\Lambda_0}^{\partial X_0}
(S_f(k)-S_{f,\infty,\Lambda}(k))P_kf\|_{L^2(\partial X_0)}
 \leq C \max_{\sigma^2_\lambda >\Lambda}\left( |k_{\lambda}|
\exp(-\epsilon '   \Im k_{\lambda})\right) \|
f\|_{L^2(\partial X_0)}.
$$
In particular, by Lemma \ref{l:mero} this holds for all
$\Lambda$ sufficiently large depending on $ k$.  We note
that the constants $C$ and $\epsilon '$ can be chosen independently 
of $k$ if $k$ is restricted to a fixed compact set $K$ on which both of
$k^2-H_{\eff}$ and $k^2-H_{\infty,\Lambda}$ are invertible and
for which $\Im k_{\lambda}>0$ when $\sigma_{\lambda}^2>\Lambda_0$.
\end{lemma}
We note that since $\pilambda f =f$, the $H^{3/2}$ norm of $f$ is
bounded by a $ \Lambda $-dependent multiple of the $L^2$ norm of $f$.
\begin{proof}
For $\Lambda \in [\Lambda_0,\infty) \cup \{\infty\}$, set
$u_{\Lambda}= (k^2-H_{\infty,\Lambda})^{-1} W_{\infty,\Lambda}P_k f$.
That means that $u_{\Lambda}$ satisfies
\begin{align*}
 (k^2+\Delta_{X_0})u_{\Lambda}& =0  \; \text{on $X_0^{\circ}$}\,, \ \ \
(\partial_n u_{\Lambda}- i P_k^2 \pilambda Ru_{\Lambda}) = -P_k^2 f.
\end{align*}

Choose $\chi \in C^{\infty}_c((-\epsilon/2,0 ])$ to be one in a 
neighborhood of 0.
 Since $U \subset X_0$ is a neighborhood of $\partial
X_0$
which can be identified with $(-  \epsilon, 0]_x\times Y_y$, we can
consider
$\chi=\chi(x)$
 to be defined on $X_0$ by extending it to be $0$ outside of
$U$.
For $g\in L^2(X)$, define 
$\pilambda  \chi g \in L^2(U)\subset L^2(X_0)$ via
$$(\pilambda  \chi g)(x,y)
=\sum _{\sigma_{\lambda}^2 \leq \Lambda}\varphi_{\lambda}(y)
\int_{y'\in \partial {X_0} }
\left(\chi(x)g(x,y')\varphi_{\lambda}(y')\right)\;d\vol_Y.$$
Then
\begin{equation}
\label{eq:ulambda}
u_{\Lambda}=(1-\chi)u_{\infty}+\pilambda\chi u_{\infty} 
+(k^2-H_{\infty,\Lambda})^{-1}
(k^2+\Delta_{X_0})(1-\pilambda)\chi u_{\infty}
\end{equation}
since the function on the right satisfies the same boundary conditions as
$u_{\Lambda} $ and is in the null space of $k^2+\Delta_{X_0}$.  

Note that by using $\pilambdanot f=f$
\begin{equation} \label{eq:uexp}  u_{\infty}\rest_U= 
\sum_{\sigma^2_\lambda \leq \Lambda}
(a_{\lambda}e^{-ik_{\lambda}x}+b_\lambda e^{i k_\lambda x })
\varphi_{\lambda}
+\sum _{\sigma^2_\lambda >  \Lambda} b_\lambda e^{i k_\lambda x }
\varphi_{\lambda}
\end{equation}
for some constants $a_\lambda$, $b_\lambda$, so that, using orthonormality of 
$ \varphi_\lambda$'s, 
\begin{align}\label{eq:pilambdau}
\nonumber 
\| u_{\infty} \|^2_{L^2} & \geq \|  u_\infty \rest_U\|_{L^2}^2 
\geq  \int_{- \epsilon}^0  
\sum _{\sigma^2_\lambda >\Lambda}| b_\lambda e^{i k_\lambda x }
|^2 dx \\
& =
\int_{- \epsilon}^0  
\sum _{\sigma^2_\lambda >\Lambda}| b_\lambda e^{i k_\lambda x }|^2 dx 
 = \sum _{\sigma^2_\lambda >\Lambda} |b_{\lambda}|^2 \frac{e^{2\epsilon \Im
    k_{\lambda}}-1}
{2 \Im k_{\lambda}}.
\end{align} 
Also, 
\[ (k^2+\Delta_{X_0}) (1-\pilambda) \chi u_{\infty}
 = [\partial_x^2, \chi] (1-\pilambda) \tilde \chi u_{\infty} \,, \]
where $ \tilde \chi $ has the same properties as 
$ \chi $ and  $ \tilde \chi \chi = \chi $. 
Our argument below takes advantage of the fact that the support of
$[\partial^2_x,\chi]$ is contained in $[\epsilon/2,0]$, while the
expansion (\ref{eq:uexp}) is valid for $x$ in $(-\epsilon, 0]$.
Hence, 
 \begin{align*}
\|(k^2+\Delta_{X_0}) (1-\pilambda) \chi u_{\infty}\|^2
& = \|[\partial_x^2, \chi] (1-\pilambda) \tilde \chi u_{\infty}\|^2\\ & 
\leq C \langle \epsilon^{-4} \rangle 
\int_{-\epsilon/2}^0 \sum _{\sigma^2_\lambda >\Lambda}
\langle k_\lambda \rangle ^2 | b_\lambda e^{i k_\lambda x }
|^2 dx \\
& \leq C \langle \epsilon^{-4}\rangle 
\sum _{\sigma^2_\lambda >\Lambda} \langle k_\lambda \rangle ^2 |b_{\lambda}|^2 \frac{e^{\epsilon \Im
    k_{\lambda}}-1}
{2 \Im k_{\lambda}}
\\ & 
= C \langle \epsilon^{-4}\rangle 
\sum _{\sigma^2_ \lambda >\Lambda} \langle k_\lambda \rangle ^2 |b_{\lambda}|^2 \frac{e^{\epsilon 2\Im
    k_{\lambda}}-1}
{2 \Im k_{\lambda}} \left( \frac{1}{e^{\epsilon \Im
    k_{\lambda}}+1}\right) \,.
\end{align*}
Thus 
(\ref{eq:pilambdau}) gives
$$\|(k^2+\Delta_{X_0}) (1-\pilambda) \chi u_{\infty}\|
\leq  C \langle \epsilon^{-2}\rangle \|u_{\infty} \| _{L^2}
\max_{\sigma_{\lambda}^2 > \Lambda}
(|k_\lambda| e^{-\epsilon \Im k_{\lambda}/2}).
$$
Using 
(\ref{eq:ulambda}),
the estimate 
$$\|(k^2 -H_{\infty,\Lambda})^{-1}
g\|_{H^{1}}
\leq (1+|k|) \|(k^2 -H_{\infty,\Lambda})^{-1}
g\|_{L^2}\,, $$ 
and the previous lemma, we obtain
\begin{align*}
& \| \pilambdanot R (u_{\infty}-u_{\Lambda})\|_{L^2(\partial X)} \leq 
\\ 
& \ \ \ C
 \langle \epsilon^{-2} +|k|\rangle 
\| (k^2 -H_{\infty,\Lambda})^{-1} \| \|u_{\infty} \| _{L^2}
\max_{\sigma^2_\lambda > \Lambda} 
\left(|k_\lambda|e^{-\epsilon \Im k_{\lambda}/2}\right). 
\end{align*}
Thus far each constant $C$ can be chosen independent of $k$, though of
course $\| u_{\infty}\|$ depends on $k$ in a continuous fashion 
on compact sets on which $k^2-H_{\eff}$ is invertible.
Note that $P_k^{-1}P_k \pilambda=\pilambda$ is a bounded operator.
Thus using the
expression
for $S_f$, $S_{f,\infty,\Lambda}$ and the previous 
lemma finishes the proof.
\end{proof}

\subsection{The cut-off in the interior}
We now turn our attention to the error introduced by using $\piN$.  Throughout
this section we assume that 
$\Lambda < \infty $.

Our results will use the following standard 
\begin{lemma}
\label{l:resth}
Suppose $ \widetilde X $ is a compact Riemannian  manifold without boundary
and $ \chi \in C^\infty_0 ( \RR ) $ is equal to $ 1 $ in a neighbourhood
of $ 0 $. Suppose that $ Y \subset \widetilde X $ is a smooth embedded
submanifold of codimension one. Then 
\[  \|  ( 1 - \chi ( -h^2 \Delta_{\widetilde X} ) ) u \rest_Y \|_{L^2 ( Y ) } 
\leq C \sqrt h \| u \|_{H^1 ( \widetilde X) } \,.\]
If
$ v \in H^2 ( \widetilde X \setminus Y ) \cap H^1 ( \widetilde X ) $
then 
\[  \|  ( 1 - \chi ( -h^2 \Delta_{\widetilde X} ) ) v \rest_Y \|_{L^2 ( Y ) } 
\leq C  h \| v \|_{H^2 ( \widetilde X \setminus Y ) } \,.\]
\end{lemma}
\begin{proof}
Both statements in the lemma are local. In fact, if $ P $ is another
elliptic second order operator on $ \widetilde X$ then 
for some constant $ C_P $ the calculus of semiclassical pseudodifferential
operators (see for instance \cite[Appendix E]{EZ}) shows that
\[  ( 1 - \chi ( - h^2 \Delta ) ) ( 1 - \chi ( -h^2 C_P (P + C_P)  ) ) = 
 ( 1 - \chi ( - h^2 \Delta ) ) + {\mathcal O}_{ H^{-k} \rightarrow 
H^{k }}  ( h^N  ) \,, \]
for all $ N $ and $ k $. Hence we can use any other second 
order elliptic operator and that property is invariant under
changes of coordinates.

It follows that  we can assume that 
$ \widetilde X = \RR^n $ and $ Y = \{ x_1 = 0 \} $, $ \RR^n \ni x = (x_1, x') $
(the compactness is irrelevant for the local statement).

Denoting the Fourier transform by $ {\mathcal F} $ we write
\begin{equation}
\label{eq:Fm} {\mathcal F}_{x' \mapsto \xi' } \left( 
 ( 1 - \chi ( -h^2 \Delta_{\widetilde X} ) ) u \rest_Y \right) 
( \xi' ) = \int_\RR ( 1 - \chi( h^2 |\xi|^2 ) ) \hat u ( \xi_1 , \xi' ) 
d\xi_1 \,. 
\end{equation}
Hence, by the Cauchy-Schwartz inequality, 
\[ 
 \|  ( 1 - \chi ( -h^2 \Delta_{\widetilde X
} ) ) u \rest_Y \|_{L^2 ( Y ) } ^2
\leq C \int_{\RR^n } F ( \xi' , h )  ( 1 - \chi( h^2 |\xi|^2 ) ) 
| \hat u ( \xi ) |^2 ( 1 + | \xi|^2 ) d \xi \,, \]
where 
\[ \begin{split}  F ( \xi' , h ) & \defeq  
\int_\RR  ( 1 - \chi( h^2 |\xi|^2 ) ) ( 1 + |\xi|^2)^{-1} d\xi_1 \\
& \leq \int_{ |\xi_1| > c/h } ( 1 + |\xi_1|^2 )^{-1} d\xi_1 +
\bbbone_{ |\xi'| > c/ h } ( \xi') \int_{\RR} ( |\xi'|^2 + |\xi_1|^2 )^{-1} d\xi_1\\
& \leq C h \,.
\end{split}  \]
This proves the first part of the lemma.

For the second part, we can 
assume that 
$ \supp v \subset \{ x \in \RR^n : |x| \leq R \} $
as we can localize to a compact set. 
We then write
\begin{equation}
\label{eq:uhat} \hat v ( \xi ) 
= \int_0^R \left( e^{-ix_1 \xi_1 } 
 {\mathcal F}_{x' \mapsto \xi' } v ( x_1 , \xi' ) 
+ e^{ i x_1 \xi_1 }  {\mathcal F}_{x' \mapsto \xi' } v ( - x_1 , \xi' ) 
\right) d x_1  \,.  \end{equation}
Since $ v \in H^1 ( \RR^n ) $, 
$ {\mathcal F}_{x' \mapsto \xi' } v ( 0 , \xi' )  \in L^2 ( \RR^{n-1} ) $ 
is well defined and 
an hence we can integrate by parts to obtain
\[ \hat v ( \xi ) = \frac 1 { \xi_1^2} 
\sum_{\pm} \left( \mp  {\mathcal F}_{x' \mapsto \xi' } \partial_{x_1} 
v ( 0\pm   , \xi' ) - \int_0 ^R 
 e^{ \mp ix_1 \xi_1 } 
 {\mathcal F}_{x' \mapsto \xi' } \partial_{x_1}^2 v ( \pm x_1 , \xi' ) 
 d x_1  \right) \,.
 \]
Since $ v \in H^2 ( \RR^n_\pm ) $, $ \partial_{x_1} v ( 0 \pm , \xi' ) $
is well defined in $ L^2 ( \RR^{n-1} )$. 
We now use the following decomposition:
\[ ( 1 - \chi ( h^2 \xi^2 )) \hat v = \hat v_1 + \hat v_2 \,, \ \ 
\hat v_1 ( \xi ) \defeq 
\bbbone_{|\xi_1| > c/h }  ( \xi) 
( 1 - \chi ( h^2 \xi^2 ))  \hat v ( \xi) \,,\]
noting that $ |\xi'| > c / h $ on the support of $ \hat v_2 ( \xi ) $.
We first estimate the contribution of $ v_2 $ 
as in the proof of the first part of the lemma:
\[ \begin{split} 
\|  v_2 \rest_Y \|_{L^2 ( Y ) } ^2
& \leq C \int_{\RR^n } G ( \xi' , h ) 
| \hat v   ( \xi ) |^2 ( 1 + | \xi'|^2 )^2 d \xi \\
& \leq \max_{\xi' \in \RR^{n-1} } G ( \xi' , h ) 
\| v \|_{ H^2 ( \widetilde X \setminus Y ) }^2 
\end{split}
\,, \]
where 
\[ \begin{split}  G ( \xi' , h ) & \defeq  
\int_{|\xi_1| < c/h } 
  ( 1 - \chi( h^2 |\xi|^2 ) ) ( 1 + |\xi'|^2)^{-2} d\xi_1 \\
& \leq \bbbone_{ |\xi'| > c/ h } ( \xi') 
\int_0^{2 c/h}  ( 1 + |\xi'|^2 )^{-2} d\xi_1\\
& \leq C h^3 \,,
\end{split}  \]
which is a better estimate than needed. 

To estimate the contribution of $ v_1 $ we use \eqref{eq:uhat}:
\[ \|  v_1 \rest_Y \|_{L^2 ( Y ) } ^2 
\leq C_R \| v \|_{H^2 ( \widetilde X \setminus Y ) } ^2 
\left( \int_{ | \xi_1 |> 1/h } \frac 1 {\xi_1^2 } d \xi_1 \right)^2 
\leq C_R h^2 \| v \|_{H^2 ( \widetilde X \setminus Y ) } ^2  \,,
\]
which completes the proof.
\end{proof}

Like $H_{\eff}$, $H_{N,\Lambda}=H_{N,\Lambda}(k)$ is a well-defined
operator
for $k\in \rsurface$.
\begin{lemma}
\label{l:resN}
Fix $\Lambda <\infty$, and suppose that $k^2-H_{\infty,\Lambda}$ is 
invertible, $k\in \rsurface$.  
Then, for $N$ sufficiently large, $k^2-H_{N,\Lambda}$ is 
invertible, and
$$\| (k^2-H_{\infty,\Lambda})^{-1} - (k^2-H_{N,\Lambda})^{-1} \|_{H^{-1}(X_0)
\rightarrow H^1(X_0)} \leq C N^{-1/4}.$$
The constant $C$ can be chosen uniformly for $k$ in a compact set
$K\subset \rsurface$ on
which $k^2-H_{\infty,\Lambda}$ is invertible.
\end{lemma}
\begin{proof}
As in \S \ref{form} we will use quadratic forms to reinterpret 
our operators.
Thus, for $N\in \Real_+$, $E\in \Complex$, 
set
\begin{align*}
q_{\infty, \Lambda}(k,E) (u,v)&  = \int _{X_0} \nabla u \overline {\nabla v}
-i \int _{ \partial X_0} W_{\infty, \Lambda}^t u 
\overline {W^*_{\infty,\Lambda} v}
- E\int _{X_0} u \overline{v} \\
&  = \int _{X_0} \nabla u \overline {\nabla v}
-i \int _{ \partial X_0} P_k \pilambda Ru \overline {P_k^* \pilambda R v}
- E\int _{X_0} u \overline{v} 
\end{align*}
and 
\begin{align*}
q_{N,\Lambda}(k,E)(u,v) & = \int _{X_0} \nabla u \overline {\nabla v}
-i \int _{  \partial X_0}  W_{N,\Lambda}^t  u 
\overline {W_{N,\Lambda}^* v}
- E\int _{X_0} u \overline{v}\\ 
&=\int _{X_0} \nabla u \overline {\nabla v}
-i \int _{  \partial X_0} P_k \pilambda R \piN  u 
\overline {P_k^* \pilambda R \piN v}
- E\int _{X_0} u \overline{v} .
\end{align*}
Here we take both form domains to be $H^1(X_0)$.  
The quadratic forms
$q_{\infty,\Lambda}(k,E)  $ and  $q_{N,\Lambda}(k,E)$  are associated to 
operators $H_{\infty,\Lambda}-E$ and 
 $H_{N,\Lambda}-E$ respectively. 
We expand the difference of the quadratic forms as follows
\begin{align*} & 
 q_{\infty,\Lambda}(k,E)(u,u) - q_{N,\Lambda}(k,E)(u,u) \\ &   
= \int _{ \partial X_0} \left ( |P_k \pilambda R u|^2
- | P_k\pilambda R \piN  u|^2 \right) \\ & 
= \
\int_{\partial X_0} \left( (P_k\pilambda Ru) \overline{P_k\pilambda R(I-\piN)u}
+  P_k \pilambda R(I-\piN)u \overline {P_k\pilambda R 
\piN u} \right) \,. 
\end{align*}

We have the following estimates:
\begin{gather}
\label{eq:est1}
\begin{gathered}
\| \Pi_\Lambda^{\partial X_0 } R ( I  - \Pi_N^{\rm{in}} ) u \|_{H^{\ell} ( \partial X_0 ) } 
\leq C_{\ell, \Lambda} \| R ( I  - \Pi_N^{\rm{in}} ) u \|_{L^2 ( \partial X_0 ) }
\leq C_{\ell, \Lambda }  N^{-1/4} \| u \|_{H^1( X_0) } 
\\
\|  \Pi_\Lambda^{\partial X_0 } R u \|_{H^{\ell} ( \partial X_0 ) }  
\leq C_{\ell, \Lambda} \| R u \|_{L^2 ( \partial X_0 ) }\leq
C_{\ell, \Lambda}  \| u \|_{H^1 ( X_0 ) } \,. 
\end{gathered}
\end{gather}
To obtain the first, we apply Lemma \ref{l:resth} to 
$ \widetilde X \defeq X_0 \sqcup X_0^\circ $ where the 
metric on $ \tilde X $ is obtained by reflecting the metric on $ X_0 $
through $ Y = \partial X_0 $. Since the metric has product structure 
near $ Y $ this means that
\[   H^1 ( X_0 ) \simeq H^1_{\rm{ev}} (\widetilde X) \,, \]
(here $ {\rm{ev}} $ refers to even functions)
and the action of the Neumann Laplacian on $X_0$ 
 is the same as the action of
$ \Delta_{\tilde X} $ on even functions. Applying Lemma \ref{l:resth} with 
$ h = 1/\sqrt N $ gives \eqref{eq:est1}.

Applying \eqref{eq:est1} to estimate the difference of the quadratic
forms we obtain, for $ E \ll 0 $, 
\[  \begin{split}
| q_{\infty,\Lambda}(k,E)(u,u) - q_{N,\Lambda}(k,E)(u,u) | &  
\leq C_{\Lambda }(k) N^{-1/4} \| u \|_{H_1}^2 
\\\ & 
\leq 
C_{\Lambda}(k)N^{-1/4}\Re (q_{\infty,\Lambda}(k,E)(u,u)) \,.
\end{split} \]
The constant depends continuously on $k$.
Here we use the fact that $\Im k_{\lambda}>0$ for all but finitely many 
$\lambda$, ensuring that $\Re q_{\infty,\Lambda}(k,E)(u,u)$ bounds 
$\|u\|_{H^1(X_0)}^2$ from above for $E \ll 0$.
Thus, by \cite[Theorem 3.4]{kato}, 
$$\| (H_{\infty,\Lambda}-E)^{-1} - (H_{N,\Lambda}-E)^{-1} \|
_{L^2(X_0)\rightarrow L^2(X_0)}
\leq C N^{-1/4} $$
for $N$ sufficiently large depending on $E$, $k$, and $\Lambda$.
This dependence on $k$ is continuous on regions where $k^2-H_{\infty,\Lambda}$
is invertible.
To extend this to other values of $E$ (in particular, $E=k^2$), we use
\begin{multline}\label{eq:resolvent}
(A-z)^{-1}=  \left\{ 
I -\left(I+(z-z_0)(B-z)^{-1}\right)(z-z_0)\left((A-z_0)^{-1}-(B-z_0)^{-1}
\right)\right\}^{-1}\\
\hspace{40mm} \times
\left(I+(z-z_0)(B-z)^{-1}\right)(A-z_0)^{-1}.
\end{multline}
Consequently, 
if $k^2-H_{\infty, \Lambda}$ is invertible, so is $k^2-H_{N,\Lambda}$
for sufficiently large $N$, with 
$$\| (k^2-H_{\infty,\Lambda})^{-1}-(k^2-H_{N,\Lambda})^{-1}\|
_{L^2(X_0)\rightarrow L^2(X_0)}
\leq C N^{-1/4}$$
Here the constant
will depend on $k$ and $\Lambda$, as will the lower bound on the $N$ for 
which this holds.  These can be chosen uniformly if $k$ is restricted to 
lie in $K$.

Now we show that there is a similar bound from $H^{-1}(X_0)$ to $H^1(X_0)$.
We choose $E$ so that 
$\Re\left(q_{\infty,\Lambda}(k,E)(u,u)\right)\geq c_0 \|u\|_{H^1(X_0)}^2 $ for some 
$c_0>0$ and all $u\in H^1(X_0)$.  Suppose 
$w\in L^2(X_0)$ and set $u=(H_{\infty,\Lambda}-E)^{-1}w$,
$u_N= (H_{N,\Lambda}-E)^{-1}w$.  
Then 
$$c_0\|u\|^2_{H^1}\leq \Re \left( q_{\infty,\Lambda}(k,E)(u,u)\right) 
= \Re (w,u) $$
so that $$c_0 \|u\|_{H^1} \leq \|w\|_{H^{-1}}.$$  This shows that we 
can (uniquely) continuously extend $(H_{\infty,\Lambda}-E)^{-1}$ to be 
a bounded operator from $H^{-1}(X_0)$ to $H^1(X_0)$ when $E\ll 0$ (the 
duality argument shows that we can extend the operator to the dual 
of $ H^1 ( X_0 ) $ and $ H^{-1} ( X_0) $ is contained in that dual as
the space of elements of $ H^{-1} ( X) $ supported in $ X_0 $).  The 
resolvent equation extends this to other values of $E$.
Likewise,
\begin{align*}
c_0 \|u-u_N\|_{H^1}^2 & 
\leq \Re \left( q_{\infty,\Lambda}(k,E)(u-u_N,u-u_N)\right) \\
 & = \Re\left ( q_{\infty,\Lambda}(k,E)(u,u-u_N)-
q_{\infty,\Lambda}(k,E)(u_N,u-u_N)\right)\\
 & = \Re\left( (w,u-u_N)-(w,u-u_N)\right. \\
& \ \ \ \ \ 
+ \left. q_{N,\Lambda}(k,E)(u_N,u-u_N) -  q_{\infty,\Lambda}(k,E)(u_N,u-u_N)\right)\\
& \leq C N^{-1/4}\|u_N\|_{H^1} \|u-u_N\|_{H^1} \\ 
& \leq C N^{-1/4}(\|u\|_{H^1}+\|u-u_N\|_{H^1} )\|u-u_N\|_{H^1}.
\end{align*}
We allow the constant $C$ to change from line to line.
This implies that 
$$\|u-u_N\|_{H^1}\leq CN^{-1/4}(\|u\|_{H^1}+\|u-u_N\|_{H^1}),$$
which then means that for sufficiently large $N$
$$\|u-u_N\|_{H^1}\leq CN^{-1/4}\|u\|_{H^1}.$$
Using (\ref{eq:resolvent}) this can be extended to other values of $E$. 
Again, these constants can be chosen uniformly for $k\in K$.
\end{proof} 

\begin{lemma}\label{l:intapprox}
Fix $\Lambda<\infty$ and $k$ so that $k^2-H_{\infty,\Lambda}$ is
  invertible
and $\Im k_{\lambda}>0$ if $\sigma^2_\lambda >\Lambda$.  
Suppose $f\in L^2(\partial X_0)$ satisfies 
$\pilambda f= f$.  Then, for $N$ so that $k^2-H_{N,\Lambda}$ is invertible,
there is a constant $C$ depending on $\Lambda$ and $k$ so that 
$$\|\pilambda P_k^{-1}(S_{f,\infty,\Lambda}(k)-S_{f,N,\Lambda}(k))P_kf\| \leq C
N^{-1/2}
\|f\|_{L^2(\partial X_0)}.
$$
The constant $C$ can be chosen independently of $k$, if $k$ is restricted
to a compact set $K\subset \Lambda_{\sigma(\Delta_{\partial X_0} )}$
on which $k^2-H_{\infty,\Lambda}$ and $k^2-H_{N,\Lambda}$ are invertible.
\end{lemma}
\begin{proof}
Choosing $N$ so that $k^2-H_{N,\Lambda}$ is invertible, set
\begin{align*}
u_{\infty}& = (k^2-H_{\infty,\Lambda})^{-1}W_{\infty,\Lambda}P_kf\; \text{and}\\
u_{N}& = (k^2-H_{N,\Lambda})^{-1}W_{N,\Lambda}P_kf.
\end{align*}
That is, $u_{\infty}$ satisfies
\begin{align*}(k^2+\Delta_{X_0})u_{\infty} &= 0 \; \text{on}\;\; X_0^{\circ}\\
\partial_n u_{\infty} -iP^2_k\pilambda R u_{\infty} & = -P_k^2 f
\end{align*}
and $u_{N}$ satisfies, for $N<\infty$,
\begin{align*}
(k^2+\Delta_{X_0}+iW_{N,\Lambda}W_{N,\Lambda}^t))u_N& =
 W_{N,\Lambda}P_kf \; \text{on}\;\; 
X_0^{\circ}\\
\partial_nu_N\restrict_{\partial X_0}& =0.
\end{align*}
Note our assumptions on $f$ mean that the $H^{3/2}$ norm of $f$ is
bounded by a $\Lambda$ dependent constant times the $L^2$ norm of
$f$.

We wish to understand $\piN u_{\infty}$.  
Let $\Psi_n$ be a real eigenfunction of the 
Neumann Laplacian on $X_0$, with $-\Delta_{X_0}\Psi_n=\tau_n^2\Psi_n$.  
Suppose in addition that $\|\Psi_n\|_{L^2(X_0)}=1$.
Then 
\begin{align*}
(\tau_n^2-k^2)(u_{\infty},\Psi_n)_{L^2(X_0)}& = 
-\int_{X_0} (\Delta_{X_0}\Psi_n u_{\infty}- \Psi_n 
\Delta_{X_0} u_{\infty})\;d\vol_{X_0}\\ 
& =\int_{\partial X_0} \Psi_n \partial_n u_{\infty}\; d\vol_Y \\
& =\int_{\partial X_0} \Psi_n (iP^2_k\pilambda R u_{\infty}-P_k^2 f)\; d\vol_Y.
\end{align*}
That is,
$$(-\Delta_{X_0}-k^2)\piN u_{\infty} 
= iW_{N,\Lambda}W_{\infty,\Lambda}^t u_{\infty}
-W_{N,\Lambda} P_k f.$$
Thus
\begin{align}\label{eq:uN}
\nonumber 
u_{N}&= \piN u_{\infty} 
- (k^2-H_{N,\Lambda})^{-1}\left( (k^2-H_{N,\Lambda})\piN u_{\infty}
- W_{N,\Lambda} P_k f \right) \\ 
\nonumber 
& = \piN u_{\infty} + i(k^2-H_{N,\Lambda})^{-1}W_{N,\Lambda}(W_{\infty,\Lambda}^t-W_{N,\Lambda}^t) u_{\infty}\\ 
& = \piN u_{\infty} + i(k^2-H_{N,\Lambda})^{-1}W_{N,\Lambda}P_k \pilambda
R (I-\piN)u_{\infty}.
\end{align}
The second part of Lemma \ref{l:resth}  gives the following estimate:
 $$\|  R (1-\Pi_N^{\rm{in} } )  u_{\infty} \|_{L^2(\partial X ) } \leq C N^{-1/2}
\| u_{\infty}\|_{H^2 (X_0 ) }\,, $$
and consequently,
$$\| P_k \pilambda R (I-\piN)u_{\infty}\| \leq C N^{-1/2}
\|u_{\infty}\|_{H^{2}(X_0)} \,$$
with constant $C$ depending continuously on $k$.

Let $g\in L^2(\partial X_0)$, $h\in H^{1/2+}(X_0)$.  Then 
$ R h \in L^2 ( \partial X_0 ) $, and 
$$|(W_{N,\Lambda}g,h)_{X_0}| = |(g, P_k^* \pilambda R \piN h)_{\partial X_0}|
\leq \| g\|_{L^2(\partial X_0)} \| h\|_{ H^{1/2^+}(X_0)}.$$
That is, 
$$\|W_{N,\Lambda}g\|_{H^{-1}(X_0)}  \leq \|W_{N,\Lambda}g\|_{H^{-1/2-}(X_0)}  \leq C \|g\|_{L^2(X_0)}$$
and the constant is independent of $N$
and depends continuously on $k$.  On the other hand, Lemma 
\ref{l:resN} shows that 
for $N$ 
sufficiently large
$$\| (k^2-H_{N,\Lambda})^{-1}\|_{H^{-1}(X_0)\rightarrow H^1(X_0)} 
\leq C. $$  Using these estimates in (\ref{eq:uN}), we find that 
$$\| u_{\infty} - u_N\|_{H^{1}} \leq C N^{-1/2 }, $$
implying the desired bound by restricting to $\partial X_0$ and 
using again the fact that $P_k ^{-1}P_k \pilambda=\pilambda$ 
is a bounded operator.
\end{proof}

\section{Proofs of Theorems}

Our proof of the Theorem in \S \ref{int}  will use the unitarity for $k$ real not
only
of the finite-dimensional scattering matrix
defined by (\ref{eq:sfinite}), but also of the approximations of the 
scattering matrix
obtained
by introducing the projections $\piN$ and $\pilambda$.
\begin{lemma}\label{l:unitary}
 Let $k\in \Real$.  Then $S(k)$ defined by
  (\ref{eq:sfinite})
and $\Pi^{\partial X_0}_{k^2}S_{f,N,\Lambda}(k)\Pi^{\partial
  X_0}_{k^2}$, for $\Lambda,\; N\in \Real_{+}\cup \{\infty\}$, are
unitary.
\end{lemma}
\begin{proof}
That $S(k)$ is unitary for $k$ real is well known.
It can be seen as follows. Recall that
$S(k)= \Pi^{\partial X_0}_{k^2}S_{f,\infty,\infty}(k)\Pi^{\partial
  X_0}_{k^2}$.

We note that $(P_k^2)^*\varphi_\lambda =
\overline{k_\lambda}\varphi_\lambda$ and $k_\lambda$ is real for
$\sigma_\lambda^2\leq k^2 $ and pure imaginary for $\sigma_{\lambda}^2
> k^2$.
Therefore
\begin{equation}\label{eq:piw}
\Pi_{k^2}^{\partial X_0} W^t(k) = \Pi_{k^2}^{\partial X_0} W^*(k)
\; \text{and $(I-\Pi_{k^2}^{\partial X_0})W^t(k)=
-(I-\Pi_{k^2}^{\partial X_0})W^*(k)$}.
\end{equation} 
Thus, we have
\begin{equation}
W(k)^{t*}\pilambda W^*(k) + W(k)\pilambda W^t(k)= 
2W(k)\pilambda \Pi^{\partial X_0}_{k^2} W^t(k).
\end{equation}
Using this and the resolvent identity gives
\begin{equation}\label{eq:rdiff}
\begin{split} & 
(k^2-H_{N,\Lambda})^{-1}- (k^2-H_{N,\Lambda}^*)^{-1}\\ &  =
i(k^2-H_{N,\Lambda})^{-1}(-\piN W^{t*}\pilambda W^*\piN
-\piN W\pilambda W^t \piN) (k^2-H_{N,\Lambda}^*)^{-1}\\
& = -2i (k^2-H_{N,\Lambda})^{-1} \piN W \Pi_{k^2}^{\partial X_0}
  \pilambda W^t \piN (k^2-H_{N,\Lambda}^*)^{-1}.
\end{split}
\end{equation}

Therefore,
\begin{equation}
\begin{split}
& \Pi^{\partial X_0}_{k^2} S_{f,N,\Lambda}(k) \Pi^{\partial X_0}_{k^2}(
\Pi_{k^2}^{\partial X_0} S_{f,N,\Lambda}(k) \Pi_{k^2}^{\partial X_0})^*\\
& = \Pi^{\partial X_0}_{k^2}\Big( I -2i
  W_{N,\Lambda}^t(k)(k^2-H_{N,\Lambda})^{-1}W_{N,\Lambda}(k)\Big) \\ &
\hspace{20mm}
\times 
\Pi_{k^2}^{\partial X_0} \Big( I +2i W_{N,\Lambda}^*(k) 
((k^2-H_{N,\Lambda})^{-1})^*
(W_{N\lambda}^t(k))^*\Big)
\Pi^{\partial X_0}_{k^2}\\
&= \Pi^{\partial X_0}_{k^2}\Big( I - 2i W_{N,\Lambda}^t(k)
\left[
  (k^2-H_{N,\Lambda})^{-1}-(k^2-H_{N,\Lambda}^*)^{-1}\right]W_{N,\Lambda}(k)
\\ & \hspace{20mm}
+4W_{N,\Lambda}^t(k^2-H_{N,\Lambda})^{-1} W_{N,\Lambda}\Pi_{k^2}^{\partial X_0}
W_{N,\Lambda}^t(k)(k^2-H_{N,\Lambda}^*)^{-1}W_{N,\Lambda}(k)
\Big) \Pi_{k^2}^{\partial X_0}
\end{split}
\end{equation}
where we have used (\ref{eq:piw}).  Applying the identity 
(\ref{eq:rdiff}) we find that 
\begin{equation}
\begin{split}
 \Pi^{\partial X_0}_{k^2} S_{f,N,\Lambda}(k) \Pi^{\partial X_0}_{k^2}(
\Pi_{k^2}^{\partial X_0} S_{f,N,\Lambda}(k) \Pi_{k^2}^{\partial
  X_0})^* = I
\end{split}
\end{equation}
as desired.
\end{proof}

\begin{theorem}\label{thm:invertible}
Let $X$ be a manifold with infinite cylindrical ends, and 
$S_{\lambda\lambda'}(k)$, $S_{f,N,\Lambda}(k)$ be as 
defined via (\ref{eq:Phi1}), (\ref{eq:Phi2}) and (\ref{eq:sfnlambda}).  
Suppose $k\in \rsurface$ and $\Lambda_0\in \Real$ are  
such that $k^2-H_{\eff}= k^2-H_{\eff}(k) $
is invertible, and $\Im k_\lambda>0$ if $\sigma_\lambda^2 > \Lambda_0$.
Then, for $\sigma_\lambda^2,\; \sigma_{\lambda'}^2 \leq \Lambda_0$ 
and $\Lambda \geq \Lambda_0,$
$$\frac{\sqrt{k_{\lambda'}}}{\sqrt{k_\lambda}}S_{\lambda \lambda'}(k)=
\langle  P_k^{-1} S_{f,N,\Lambda}(k) P_k \varphi_{\lambda'}
, \varphi_\lambda  \rangle 
+{\mathcal O}( N^{-\frac12} +  
e^{- \Lambda/C } ) \,.$$
\end{theorem}
 We recall that $k^2-H_{\eff}$ is
invertible if $k$ is in the physical space with $\Im k>0, \; \Im
k_{\lambda}>0$ for all $\lambda$, and that 
$ ( k^2 - H_{\eff})^{-1} $ is meromorphic on $ \Lambda_{\sigma(\Delta_{\partial
X_0 } )} $.

\begin{proof}
The proof follows from writing
\begin{equation}\label{eq:approx}
\begin{split}
\frac{\sqrt{k_{\lambda'}}}{\sqrt{k_\lambda}}S_{\lambda \lambda'}(k)&  =
\langle  P_k^{-1}S_{f}(k)P_k\varphi_{\lambda'}, \varphi_{\lambda}\rangle\\
& = \langle  P_k^{-1} S_{f,N,\Lambda}(k)P_k\varphi_{\lambda'},
\varphi_\lambda
\rangle
+
\langle 
P_k^{-1}(S_{f}(k)-S_{f,\infty,\Lambda}(k))P_k\varphi_{\lambda'},
\varphi_{\lambda}\rangle 
\\ & \hspace{20mm}+ \langle 
P_k^{-1}(S_{f,\infty,\Lambda}(k)-S_{f,N,\Lambda}(k))P_k
\varphi_{\lambda'},\varphi_{\lambda}\rangle \,, 
\end{split}
\end{equation}
where we note the first equality follows from Proposition \ref{prop1}.
Applying
Lemma \ref{l:23} and Lemma \ref{l:intapprox}, we obtain the 
theorem.
\end{proof}

We now prove Theorem \ref{theorem:intro}.
\begin{proof} 
If $k^2-H_{\eff}$ is invertible this is just Theorem \ref{thm:invertible},
and hence 
it remains to prove that the estimate is valid for all $k \in \Real$,
even 
if $k^2-H_{\eff}$ is not invertible. 

Using
the 
unitarity proved in Lemma \ref{l:unitary}, along with the fact that
$\sigma_{\lambda'}^2\leq k^2$, $\sigma_\lambda^2 \leq k^2$,
 we see that each of the
  terms on the right hand side of (\ref{eq:approx}) is bounded for all
  $k
\in \Real$. Also, 
for $N,\; \Lambda \in \Real_+\cup \{\infty\}$
$(\sqrt{k_{\lambda'}}/\sqrt{k_{\lambda}})\langle  
S_{f,N,\Lambda}(k)\varphi_\lambda, \varphi_{\lambda'} \rangle$ 
has a meromorphic extension to 
$\Lambda_{\sigma(\Delta_{\partial X_0} )}$, as 
can be seen from the formula \eqref{eq:sfnlambda} and the fact that 
$ ( k^2 - H_{N, \Lambda })^{-1} $ continues meromorphically to 
$ \Lambda_{ \sigma ( \Delta_{ \partial X_0} )} $.
Hence 
\[ k\in  ( - \sigma_\lambda , \sigma_\lambda ) \cap ( - \sigma_{\lambda'}
, \sigma_{\lambda'} ) \,, \]
has a neighborhood in $\Lambda_{\sigma(\Delta_{\partial X_0} )}$ on which 
$\langle  S_{f,N,\Lambda}\varphi_{\lambda'},\varphi_\lambda \rangle$ 
is holomorphic,
$N, \Lambda \in \Real_+\sqcup\{\infty\}$.

We will now apply the maximum principle:
(\ref{eq:approx}) and
Lemmas \ref{l:23} and \ref{l:intapprox} show that 
\[  S_{\lambda \lambda'}  - 
 \langle S_{f,N,\Lambda}\varphi_{\lambda'}, \varphi_\lambda \rangle \,,\]
is bounded by $ C ( N^{-\frac12} + e^{-\Lambda/C} ) $ on the boundary 
of the neighbourhood chosen above, since $ ( k^2 - H_{\eff} )^{-1} $
is bounded there. The theorem follows as the difference is
holomorphic. 

In other words,
we have shown the theorem holds when $k\in \Real$ is on the boundary of
the physical space even if 
$ k $ is a pole of $(k^2-H_{\eff})^{-1}$, as long as $k^2
\not = \sigma_\lambda^2, \sigma_{\lambda'}^2$.

To finish the proof, consider what happens at a point $k_0 \in \Real$,
$k_0^2=\sigma_{\lambda}^2\geq \sigma_{\lambda'}^2$.
(The case $\sigma_{\lambda}^2<\sigma_{\lambda'}^2$ follows by symmetry.)
If $\sigma_{\lambda}^2=\sigma_{\lambda'}^2$, then 
since $\sqrt{k_{\lambda'}}/\sqrt{k_{\lambda}}=1$ (except for the removeable
singularity
at $k_\lambda=0$),
$\langle S_{f,N,\Lambda}(k)\varphi_{\lambda'}, \varphi_\lambda \rangle$ 
has a  meromorphic extension to a neighborhood of $k_0$ in 
$\Lambda_{\sigma(\Delta_{\partial X_0} )}$.
The boundedness at
$k_0$, again obtained from unitarity, 
ensures that there exists a neighborhood of $k_0$ on which 
$\langle  S_{f,N,\Lambda}\varphi_{\lambda'}, \varphi_\lambda \rangle$
is holomorphic.  Thus the previous argument 
using the maximum principle holds here as well.

Now suppose $\sigma_{\lambda'}^2<\sigma_{\lambda}^2=k_0^2$, and set 
$T_{\lambda \lambda'}(k)= (\sqrt{k_{\lambda'}} /\sqrt{k_{\lambda}})
S_{\lambda \lambda'}(k)$.  Then $T_{\lambda \lambda'}$ is meromorphic
in a neighborhood of $k_0$.    Using the unitarity of $S(k)$ for $k$ real, 
$$|T_{\lambda\lambda'}(k)|^{2}
\leq |k_{\lambda'}||k_{\lambda}|^{-1}\; 
\text{ for $ k^2 \geq \sigma_{\lambda^2},\; k \in \Real$}.$$ 
Thus $\sqrt{k_{\lambda}}T_{\lambda \lambda'}(k)$ is bounded at $k_0$,
and $T_{\lambda \lambda'}$ must then also be bounded at $k_0$,
since near $k_0$ it is a meromorphic function of $k_\lambda$.  Therefore 
$S_{\lambda\lambda'}(k_0)=0$.  Since we have in fact only used the 
unitarity of $S(k)$ for $k\in \Real$ 
and the existence  of a meromorphic extension, the
same argument gives
$$\langle  S_{f,N,\Lambda}(k)\varphi_{\lambda'}, \varphi_\lambda
\rangle\restrict_{k=k_0}= 0;\ 
\text{for $\sigma_{\lambda'}^2<\sigma_{\lambda}^2= k_0^2$; $N,\Lambda 
\in \Real_{+}\cup \{\infty\}$ }.$$
Thus the approximation is exact in this special case.
\end{proof}

\section{An example}
\label{ex}
In this section we consider the simplest one-dimensional example
where
things are explicitly computable and we are able to see the effects of
the 
approximation $\piN$ explicitly.

\begin{figure}
\begin{center}
\includegraphics[width=6in]{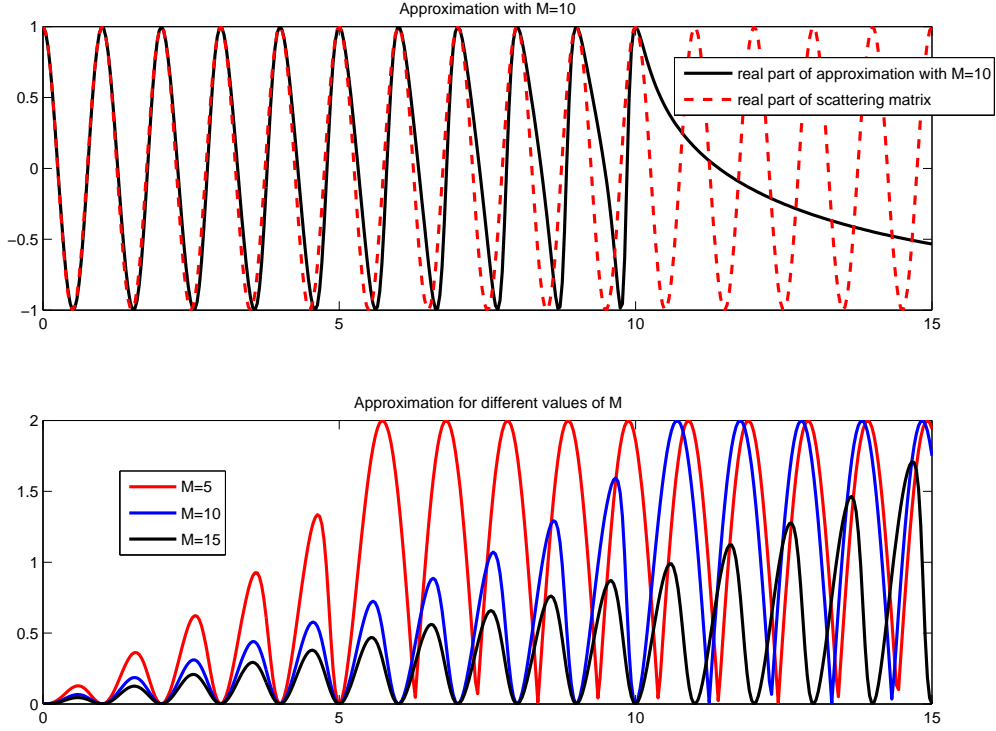}
\end{center}
\caption{An illustration of the example in \S \ref{ex}: the top figure
shows the real parts of the approximation and of the scattering matrix, and
the lower one, the graphs of $ | S_{M^2} ( k ) - S ( k ) | $ for 
different values of $ M $.}
\label{f:1}
\end{figure}

Let $X=(-\pi,\infty)$, with $X_0=(-\pi, 0]$ and $X_1=[0,\infty)$.
We consider the operator $-\partial^2_x$ on $X$, with Neumann boundary 
conditions.  Although strictly speaking this example does not fall
in the class considered in the first part of the note ($\overline{X}$
has a boundary, $\{-\pi\}$), it is easy to see the arguments of 
the previous sections follow through, with $\partial X_0$ replaced by 
$Y=\{0\}$.  Because $Y$ is 
a point, the full scattering matrix is a scalar, and
is easily computed to be $S(k)=e^{2\pi i k}$.

For this example,
$$
\Psi_n(x)=\left\{ 
\begin{array} {ll}
\pi^{-1/2} & \;\text{if $n=0$}\\
(2/\pi)^{1/2}\cos (n x)& \text{\ if $n>0$}.
\end{array}
\right.
$$
Since there is no sense in the cutoff $\pilambda$ for this
problem,
we use only one subscript on our approximations of $W$:
$$W_{M^2}(k)\defeq \sqrt{k}\sum_{n=0}^M \Psi_n(0) \Psi_n(x).$$
Similarly, we denote the approximation of $S(k)$ thus obtained by 
$S_{M^2}(k)$.
In the notation of the paper $ M = \sqrt N $.
We denote by $\widetilde{W}_{M^2}= \widetilde{W}_{M^2}(k)$ the $M+1$ vector 
$\pi^{-1/2}(1, \sqrt{2},...,\sqrt{2})^t$.

Note that if $i a^t a\not = -1$, 
$$(I+i a a^t)^{-1}= I -i (1+i a^t  a)^{-1} a a^t.$$
Set $D_{M^2}=D_{M^2}(k)$ to be the $M+1\times M+1$ matrix given by 
$((k^2-n^2)\delta_{nm})$.  We see that
when $k\not \in \Integers$ so that $D_{M^2}(k)$ is invertible,
 the approximation $S_{M^2}(k)$ is
given by 
\begin{align*} S_{M^2}(k)&=-1+2i \widetilde{W}_{M^2}^t 
(D_{M^2} +i \widetilde{W}_{M^2} \widetilde{W}_{M^2}^t)^{-1} \widetilde{W}_{M^2}\\
& = -1+ 2i (D_{M^2}^{-1/2}\widetilde{W}_{M^2})^t 
(I + i D^{-1/2}_{M^2} \widetilde{W}_{M^2}
( D^{-1/2}_{M^2}\widetilde{W}_{M^2})^t)^{-1} D_{M^2}^{-1/2}\widetilde{W}_{M^2}.
\end{align*}
Set $B_{M^2}= D_{M^2}^{-1/2}\widetilde{W}_{M^2}$ and $\beta_{M^2}= B_{M^2}^t
B_{M^2}$.
Then, for $k \not \in \Integers$,
\begin{align}\label{eq:SN}
S_{M^2}(k)& = -1+ 2 i B_{M^2}^t(I -i (1+i\beta_{M^2})^{-1}
B_{M^2}B_{M^2}^t)B_{M^2} \nonumber \\
& = -1+ 2i\left( \beta_{M^2}-i
\frac{\beta_{M^2}^2}{1+i\beta_{M^2}}\right).
\end{align}
Now 
\begin{equation}\label{eq:betaN}
\beta_{M^2}=\frac{1}{\pi}\left( \frac{1}{k}
+ \sum_{n=1}^{M} \frac{2k}{k^2-n^2}\right).
\end{equation}
We note that 
$$\lim_{M\rightarrow \infty} \beta_{M^2}= \cot \pi k;$$
one can use this and  (\ref{eq:SN}) to see that 
$$\lim_{M\rightarrow \infty} S_{M^2}(k)=e^{2\pi i k}=S(k)$$
when $k^2\not \in \Natural_0$.  
Using (\ref{eq:SN}) and (\ref{eq:betaN}), we see that for $k \in \Real
\setminus \Integers$ and $M>|k|$, 
$$C_1 /M \leq |S_{M^2}(k)-S(k)| \leq C_2/M$$
for some positive constants $C_1$, $C_2$ depending on $k$.  Since
 $ M = \sqrt N $, this shows that
the estimates obtained in Lemma \ref{l:intapprox} and in the main 
theorem are optimal.

\noindent

\end{document}